%% file: aanda.tex
\documentclass[longauth]{aa}

\usepackage{graphicx}
\usepackage{txfonts}
\usepackage{float}
\usepackage{natbib}
\usepackage{multirow}
\usepackage[switch]{lineno} 
\usepackage{xspace}
\usepackage{nccmath}
\usepackage{array}
\usepackage{arydshln}
\usepackage{stfloats} 
\usepackage[addedmarkup=bf]{changes}

\usepackage{subfigure}
\usepackage[range-units=single,range-phrase=\text{--}]{siunitx}
\usepackage{makecell}
\usepackage[colorlinks,allcolors=blue]{hyperref}
\makeatletter
\renewcommand*\aa@pageof{, page \thepage{} of \pageref*{LastPage}}
\makeatother
\newcommand{\RomanNumeralCaps}[1]
    {\MakeUppercase{\romannumeral #1}}
\begin{document} 
\newcommand{\gam}{$\gamma$\xspace}
\newcommand{\gammapy}{\textsc{Gammapy}\xspace}
\newcommand{\fermipy}{\textsc{Fermipy}\xspace}
\newcommand{\hess}{H.E.S.S.\xspace}
\newcommand{\fermilat}{\emph{Fermi}-LAT\xspace}
\newcommand{\source}{HESS~J1813$-$178\xspace}
\newcommand{\psr}{PSR~J1813$-$1749\xspace}
\newcommand{\fermi}{4FGL~J1813$-$1737e\xspace}
\newcommand\amwm[1]{{\color{red} #1}}

   \title{Unveiling extended gamma-ray emission around HESS~J1813$-$178}

   \input{authors}


   \date{March 25, 2024}

 
  \abstract
   {\source is a very-high-energy $\gamma$-ray source spatially coincident with the young and energetic pulsar \psr and thought to be associated with its pulsar wind nebula (PWN). Recently, evidence for extended high-energy emission in the vicinity of the pulsar has been revealed in the \emph{Fermi} Large Area Telescope (LAT) data. This motivates revisiting the \source region, taking advantage of improved analysis methods and an extended data set. }
  {Using data taken by the High Energy Stereoscopic System (H.E.S.S.) experiment and the \fermilat , we aim to describe the $\gamma$-ray emission in the region with a consistent model, to provide insights into its origin.}
   {We performed a likelihood-based analysis on 32 hours of H.E.S.S. data and 12 years of \fermilat data and fit a spectro-morphological model to the combined datasets. These results allowed us to develop a physical model for the origin of the observed $\gamma$-ray emission in the region.}
   {In addition to the compact very-high-energy $\gamma$-ray emission centered on the pulsar, we find a significant yet previously undetected component along the Galactic plane.  
   With \fermilat data, we confirm extended high-energy emission consistent with the position and elongation of the extended emission observed with \hess  
   These results establish a consistent description of the emission in the region from GeV energies to several tens of TeV.}
   {This study suggests that \source is associated with a $\gamma$-ray PWN powered by \psr. A possible origin of the extended emission component is inverse Compton emission from electrons and positrons that have escaped the confines of the pulsar and form a halo around the PWN.}


   \maketitle
%

\section{Introduction and multiwavelength context}
Results of the Galactic plane survey, conducted by the Imaging Atmospheric Cherenkov Telescope (IACT) Array H.E.S.S. \citep{firstHGPS, HGPS, HGPS2}, have significantly improved our knowledge of the Galactic very-high-energy (VHE) $\gamma$-ray sky. This survey detected many previously unknown sources, many of which have multi-wavelength counterparts, shedding light on the acceleration mechanism of charged particles. 
However, due to a large number of possible acceleration sites, firm identification of VHE sources often remains elusive, with high-energy $\gamma$-rays being mainly produced by either the interaction of leptons with interstellar radiation fields or inelastic collisions of hadronic cosmic rays with nuclei of the interstellar medium such as molecular clouds. Therefore, $\gamma$-rays are a valuable probe for understanding the acceleration of particles to extreme energies.

\source is a bright source discovered in 2006, located directly on the Galactic plane (GLON: $12.82\pm 0.03^\circ$, GLAT: $-0.03\pm 0.02\degr$) with a $1\,\sigma$ radius of $\sigma = (0.050 \pm 0.004)\degr$,  \citep{HGPS2}. This discovery was confirmed by the IACT system MAGIC \citep{magic}, and the detection of an X-ray counterpart was reported \citep{integral, xmm-newton, chandra}. \source is positionally coincident with the young shell-type radio supernova remnant G012.8–00.0, with an estimated age of $300-2500\,$years \citep{w33, psr_distance1} and the SNR G012.7–00.0, as well as the young, energetic pulsar \psr with a characteristic age of $5600\,$years and a spin-down luminosity of $\dot{E} = 5.6 \times 10^{37}\,\text{erg}\text{s}^{-1}$ \citep{psr, psr_distance1}.

Additionally to the compact emission, extended emission, labeled as HGPSC 063, was observed, but it was believed to be a potential background artefact and therefore discarded \citep{HGPS2}. 

In contrast to the $0.036\degr$ extended source detected at TeV, analysis of GeV \fermilat data revealed emission that is positionally coincident with \source, but with an extension of $0.6\degr \pm 0.06\degr$ \citep{fermi_araya}. A recent study of the region with \fermilat supports the detection of extended emission and estimates a spatial extent of the emission of $0.56^\circ$ in an energy range of $1\,\text{GeV} - 20\,\text{GeV}$ \citep{fermi_2021}. 
While previous studies of the TeV emission favour a leptonic origin connected to the pulsar \citep{xmm-newton}, \citet{fermi_araya} concluded that the origin of the GeV $\gamma$-ray emission is more likely to be cosmic rays accelerated at the shock fronts of the Supernova Remnant (SNR) or the H\RomanNumeralCaps 2 star-forming region W33, located at a distance of $10\arcmin$ from \source. 
In this case, cosmic rays collide with ambient gas, producing $\pi_0$ which then decay into two $\gamma$-ray photons.

Studies of the region have also been conducted by HAWC, revealing the extended emission 3HWC~J1813$-$174, with an extension of less than $0.5^\circ$ \citep{hawc_cat}. LHAASO observations of this region also reveal extended emission. The source LHAASO~J1814$-$1719u$^\ast$ observed with an extension of $(0.71 \pm 0.07)^\circ$ by the Water Cherenkov Detector Array (WCDA) and an extension of $<\,0.27^\circ$ in the one square kilometer array (KM2A) data, as well as the source LHAASO~J1814$-$1636u only observed by the KM2A array with an extension of $(0.68 \pm 0.08)^\circ$ \citep{lhaaso}. 
Note, however, that there are intrinsic uncertainties in the association of sources detected by the WCDA and KM2A LHAASO facilities, with LHAASO~J1814$-$1719u$^\ast$ indicated as an uncertain merger, such that the extended LHAASO~J1814$-$1719u$^\ast$ observed by WCDA may actually correspond to the extended LHAASO~J1814$-$1636u observed by KM2A. 

The confirmation of the detection of large, extended emission in the region around \source by the Water Cherenkov Detectors, and improvements in background rejection and event reconstruction, have recently led to the discovery of previously undetected, large extended emission in IACT data \citep{halo1, halo2}. This is a particularly interesting development for the region around \source and warrants a reanalysis of the region to resolve the disagreement between the different observed morphologies.

While pulsar halos have been detected in evolved systems, regions powered by the emission from young pulsars usually present a confined PWN \citep{pwn_population}. Therefore, a significant detection of extended $\gamma$-ray emission in the TeV energy range around \psr will help to resolve the discrepancy in extension between GeV and TeV energies. Observing the escape of electrons from the PWN into the interstellar medium at such a young age is highly unlikely, therefore the detection of extended emission resulting from highly relativistic electrons could hint at an untypical behaviour of the pulsar or an early interaction of the reverse shock of the SNR with the pulsar wind, disrupting the system and facilitating an early release of particles into the surrounding Interstellar medium (ISM).

We present an updated analysis of the H.E.S.S. data in the region around \psr. To obtain a consistent description of the region, we carry out an analysis of the \fermilat data, with increased exposure compared to the study conducted in \cite{fermi_araya}, and a joint likelihood minimisation of these datasets, allowing us to analyse the data from both experiments simultaneously and obtain a consistent description. We additionally present a leptonic and hadronic model in order to explain the emission observed around \psr. 



\section{Data analysis}

\subsection{H.E.S.S.}
\subsubsection{Data selection}
The H.E.S.S. array consists of four imaging atmospheric Cherenkov telescopes with a mirror diameter of $12\,$m, constructed in 2003, located in the Khomas Highland in Namibia \citep{sys_index}. A fifth $28\,$m diameter telescope was added to the array in 2012. The telescopes are capable of detecting photons in an energy range from a few tens of GeV up to $100$~TeV. The data used in this analysis was obtained between 2004 and 2010 as part of different observation campaigns in the sky region, therefore the fifth telescope has not been used in the following analysis.

H.E.S.S. observations consist of data collection intervals of $\sim28\,$min, called observation runs. The selection of suitable data is based on the observation characteristics for each run. For this data analysis, participation of at least three telescopes per run was required. An additonal requirement was the pointing position of the telescopes, which was required to not be offset by more than $2.0\degr$ from $\text{RA}=273.40^\circ$, $\text{Dec}=-17.84\degr$, the position of \source as derived in \citet{HGPS}.

After applying standard selection cuts used in spectral analysis \citep{sys_index}, this selection resulted in a dataset with a total exposure of 32 hours. The average angular distance of the position of \source to the centre of the field of view (FoV) in all observations of the dataset is $1.28\degr$, as most observations were taken around a neighboring source, HESS~J1809-193 (Fig. \ref{fig-appendix:exposure}, \citealt{J1809}). 

The data reduction was performed using the H.E.S.S. analysis package (see \citealt{sys_index}). The data was reconstructed using the \texttt{ImPACT} (Image pixel-wise fit for atmospheric Cherenkov Telescope) algorithm, which determines the best-fit parameters of the $\gamma$-ray-like events by using a template-based maximum-likelihood approach \citep{impact}.

An important source of background in the analysis of IACT data are atmospheric air showers initiated by charged cosmic-ray particles. Cosmic-ray showers outnumber the showers initiated by $\gamma$-rays by several orders of magnitude. 
To describe this background across the whole field of view, a template model was used. This model was constructed from a large set of runs taken in regions without known $\gamma$-ray sources, following the scheme described in \citet{fov-bkgmodel}. The flux normalisation and spectral index of this field of view background template were used used to account for the different observation conditions in each run (for more information see Appendix \ref{bkg}). 

For each individual run, a safe energy threshold was defined to exclude the threshold region where effective detection areas vary steeply with energy, and where the energy reconstruction is biased. The threshold was chosen as the energy at which the energy bias (the deviation between true and reconstructed event energy \citep{sys_index}) is below 10\%. A second energy threshold was defined as the peak in the field of view background template spectrum \citep{fov-bkgmodel}; the higher of the two thresholds was applied for the run.

The events passing the cut criteria were then stacked into a single dataset and the field of view background template, as discussed in \citet{fov-bkgmodel}, was applied. In addition to the run-wise threshold, a global energy threshold for the dataset of $0.4$~TeV was set. All events below this energy were excluded from the binned-likelihood estimation. 

\subsubsection{Data processing}
\label{eqns_fitmodel}
The open-source Python package \texttt{gammapy, version 0.18.2} \citep{gammapy_zenodo, gammapy} was used for the data analysis, to characterise the spectral and morphological properties of the $\gamma$-ray emission. \texttt{Gammapy} enables a three-dimensional likelihood analysis that stores the data and models in data cubes, allowing a simultaneous parameter optimisation of the model in the two spatial dimensions and energy. All results were cross-checked using an independent calibration and reconstruction scheme \citep{hap-fr}.

The background model flux normalisation and spectral index were fitted to each run used in the analysis to account for differences in atmospheric conditions and changes in the optical efficiency of the telescopes over time. The list of events passing the selection criteria was then combined into a data cube of $4\,^\circ\times 4\,^\circ$, centred on the previously derived position of \source in the equatorial coordinate system. This size ensures a sufficiently large off-source region for estimating the rate of remaining hadronic background events, without adding a large number of unrelated $\gamma$-ray sources.
Spatial bins of $0.02\,^\circ$ size, and a logarithmic energy binning of 25 bins between $0.1\,$TeV, and $100\,$TeV, were chosen for this study. 

A likelihood fit to determine the best description of the emission in the FoV was then conducted. For this purpose symmetric and elongated 2-dimensional Gaussian source models where used to describe the morphology. The symmetric Gaussian is defined as:

\begin{linenomath}
\begin{ceqn}
\begin{equation}
    \Phi\,(\Theta) = \frac{1}{2\pi \sigma_M^2} \exp{\left(-\frac{1}{2}\frac{\Theta^2}{\sigma_M^2}\right)}\,\,,
\end{equation}
\end{ceqn}
\end{linenomath}
with $\sigma_M$ the major axis of the Gaussian and $\Theta$ the angular distance to the model centre. In the case of an elongated Gaussian, the major axis $\sigma_M$ is defined by $\sigma_\text{eff}$, the effective containment radius:
\begin{linenomath}
\begin{ceqn}
\begin{equation}
    \sigma_\text{eff}\,(\Delta \phi) = \sqrt{\left(\sigma_M \sin{(\Delta \phi)}\right)^2 + \left(\sigma_M \cdot \sqrt{(1-e^2)}\cos{(\Delta \phi)}\right)^2}\,\,,
\end{equation}
\end{ceqn}
\end{linenomath}
with $e$ the eccentricity and $\Delta \phi$ the difference between the position angle of the Gaussian $\phi$, and the position angle of the evaluation point.
\\
As a spectral model, a power law was used. This model is described by:

\begin{linenomath}
\begin{ceqn}
\begin{equation}
    \frac{\text{dN}}{\text{dE}} = \text{N}_0 \cdot \left(\frac{\text{E}}{\text{E}_0}\right)^{-\Gamma}\,\,,
\end{equation}
\end{ceqn}
\end{linenomath}
where $\text{N}_0$ is the flux normalisation factor, $\Gamma$ the spectral index and $\text{E}_0$ the reference energy, which is set to $1\,$TeV for the whole study. Alternatively, a logarithmic parabola spectral model was used:

\begin{linenomath}
\begin{ceqn}
\begin{equation}
        \frac{\text{dN}}{\text{dE}} = \text{N}_0 \cdot \left(\frac{\text{E}}{\text{E}_0}\right)^{-\Gamma - \beta \log\left(\frac{\text{E}}{\text{E}_0}\right)}\,\,,
\end{equation}
\end{ceqn}
\end{linenomath}
where $\beta$ is the curvature, as well as a power law with a generalised exponential cut-off

\begin{linenomath}
\begin{ceqn}
\begin{equation}
        \frac{\text{dN}}{\text{dE}} = \text{N}_0 \cdot \left(\frac{\text{E}}{\text{E}_0}\right)^{-\Gamma} \cdot \exp{(-\lambda \text{E})}\,\,,
\end{equation}
\end{ceqn}
\end{linenomath}
with $\lambda$, the inverse of the cut-off energy. 

Significance maps of the region were computed using the method described in \cite{lima}. For all sky maps, a correlation radius of $0.4^\circ$ was applied, since this has proven beneficial for detecting extended emission in previous studies \citep{halo1}. The choice of correlation radius has no influence on the analysis, it only improves the visibility of large extended emission. 

\subsubsection{Estimation of systematic uncertainties}
The Instrument Response Functions (IRFs) describe the response of the system under various observation conditions. The IRFs for each run were obtained by interpolating between IRFs generated from Monte Carlo simulations covering a grid of observational conditions \citep{IRFs}. This method cannot describe all variations of observational conditions in detail, which leads to systematic uncertainties affecting the likelihood minimisation. In this analysis, three possible sources of uncertainties were considered in our results.

Firstly, a discrepancy between the true and reconstructed pointing positions will lead to uncertainties in source position estimation. As indicated in \citet{pointing}, the uncertainty of the pointing is $20\,\arcsec$, estimated using the position of known $\gamma$-ray sources and stars. 

Secondly, the reconstruction of the energy of a $\gamma$-ray strongly depends on the optical efficiency of the telescopes, influenced by effects such as degradation of the photomultiplier tubes or mirrors over time. Additionally, the transparency of the atmosphere, affected by effects such as the presence of aerosols, strongly influences the accuracy of the energy reconstruction. A discrepancy in energy reconstruction particularly influences the estimated spectral parameters of the source. 

Thirdly, an important aspect of the data analysis is the correct statistical description of the inferred background event rate as a function of energy. The background model used in this analysis was constructed using observation runs taken under not fully identical observation conditions from those runs used in this analysis. The statistical and systematic uncertainties in the inferred background model will lead to uncertainty in the prediction of background counts and affect all fit parameters.

To assess the systematic errors, a large number of pseudo-datasets with a shift of the energy scale, as well as a shift of the amplitude and index of the background model were created, and a linear gradient to the hadronic background model was applied. Then, a likelihood minimisation for these pseudo-datasets was performed. Using this method, the spread of the resulting fit values can be interpreted as the systematic error of the respective parameter. In addition, the systematic error in the pointing position is added. A more detailed description of the method can be found in Appendix \ref{appendix:sys}, and the systematic errors estimated for the source parameters are presented in Table \ref{tab-hess:fitparamters}.

\subsection{\fermilat Data}
The LAT onboard the \emph{Fermi} satellite detects $\gamma$-rays between $20\,$MeV and more than $300\,$GeV from the entire sky by pair-conversion within the detector \citep{LAT_performance}.

In this study, the data from the beginning of the mission in August 2008 until October 2021 was analysed. The most recent IRFs from Pass 8 version 3 --- \texttt{P8R3\_SOURCE\_V2}~\citep{sensitivity} were used and the catalogue models were taken from the 12-year 4FGL source catalogue~\citep{4FGL}. A $6\degr$ region of interest (ROI) around the source position of \fermi, as derived in~\citet{fermi_araya}, a bin size of $0.025\degr$ and 8 energy bins per decade with logarithmic spacing were used. Only events arriving with a maximum zenith angle of $90\degr$ were considered, in order to avoid including secondary $\gamma$-rays from the Earth's horizon.

The background was modeled using a isotropic background model (\texttt{iso\_P8R3\_SOURCE\_V3\_v1.txt}), and a Galactic diffuse background model (\texttt{gll\_iem\_v07.fits}).
In order to describe the region well, all known catalogue sources in a region of $10\degr$ around the source position were included. 

For the analysis, the Python package \texttt{fermipy, version 1.0.1}~\citep{fermipy}, optimised for performing a binned likelihood analysis of \fermilat data using the \emph{Fermi} Science Tools, was used. With \texttt{fermipy} a counts cube and IRF cubes in an energy range between $1\,$GeV and $1\,$TeV was created. 
For the analysis of the source, the data, IRFs, and background models were exported to \texttt{gammapy}. In contrast to the standard analysis chain of the \fermilat data, this allows us to fit both morphological and spectral parameters simultaneously.

\section{Results}
\subsection{H.E.S.S. Results} \label{result}
\begin{figure*}
\begin{minipage}[b]{.49\textwidth}
\includegraphics[width=\textwidth]{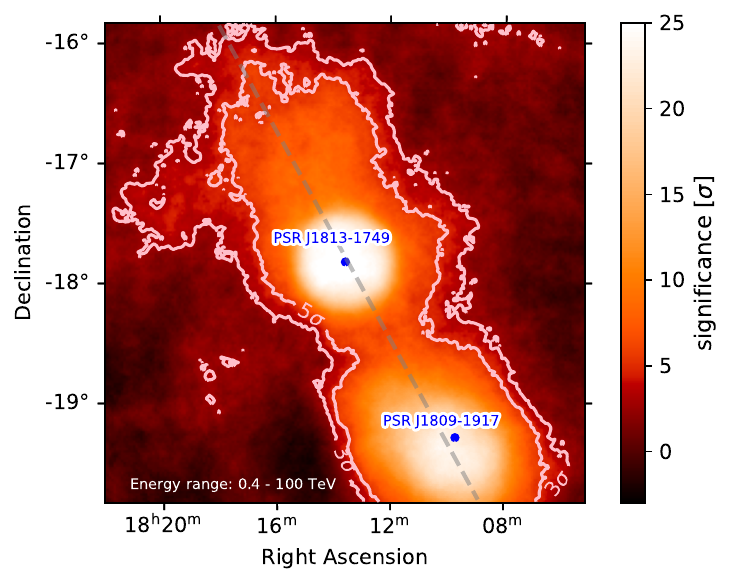}
\end{minipage}\qquad
\begin{minipage}[b]{.49\textwidth}
\includegraphics[width=\textwidth]{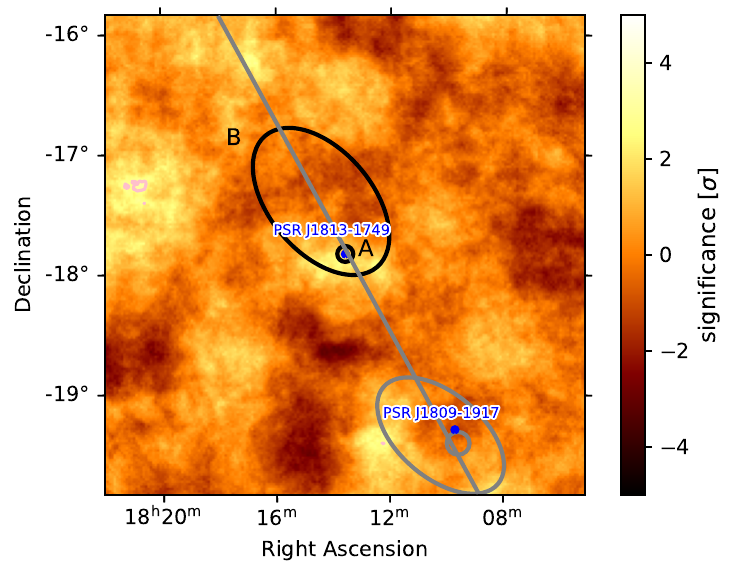}
\end{minipage}
\caption{Significance maps of the data taken by \hess in the energy range between $0.4\,$TeV to $100\,$TeV. For both maps, a correlation radius of $0.4^\circ$ was used. Left: Significance map of the region around \psr. The contours correspond to the $3\,\sigma$ and $5\,\sigma$ regions. The dashed line indicates the position of the Galactic plane. Right: Significance map of the region after subtracting the emission using four components. The blue dots indicate the positions of both pulsars. The $1\,\sigma$ Gaussian extend of the models used to describe the emission are indicated by the grey and black lines.}
\label{fig-hess:morphology}
\end{figure*}

A significance map of the region, computed with a correlation radius of $0.4^\circ$, is shown in Fig. \ref{fig-hess:morphology}.  A significance map computed with a correlation radius of $0.06^\circ$, which is equivalent to the point-spread-function (PSF) of the analysis, has also been computed and can be seen in figure \ref{fig-appendix:sig_small_corr}.
Bright $\gamma$-ray emission around the pulsar \psr is visible. Additionally, one can observe diffuse, extended emission around the pulsar, as well as $\gamma$-ray emission from the source HESS~J1809$-$193 an extended source potentially powered by the pulsar PSR~J1809$-$1917. The $\gamma$-ray emission from this source was accounted for by using the best-fit description derived in \citet{J1809}. The $1\,\sigma$ extends of the two Gaussian models used in the analysis are indicated by the grey dashed lines in the right panel of Fig. \ref{fig-hess:morphology}.

The compact emission can be accounted for using a Gaussian model, centred at a position of $\text{R.A.}=(273.396 \pm 0.004)\degr, \text{Dec.}=(-17.831 \pm 0.004)\degr$ and  $\sigma =(0.056 \pm 0.003)\degr$, hereafter referred to as HESS~J1813$-$178A. The emission is well described with a power-law spectral model and detected with a significance of $38\,\sigma$. The best-fit position and extension derived in this analysis are compatible with the results reported in \citet{HGPS}. 

In addition to the compact emission centred at the location of \psr, a second extended emission component was detected during the Galactic plane survey \citep{HGPS2}, but discarded because of large uncertainties in the background estimation. The emission was reported to be located at $\text{R.A.}=(273.46 \pm 0.05)\degr, \text{Dec.}=(-17.77 \pm 0.06)\degr$ with a Gaussian width of $\sigma=(0.31 \pm 0.06)\degr$ \citep{HGPS2}. In addition to the improved image reconstruction technique ImPACT \citep{impact}, this study used a background template created from a large number of archival observations in order to estimate the background events, as opposed to the former approach, which only used a ring-shaped area around the source from the observation run itself and is therefore heavily biased by the initial assumption of an extension of the emission \citep{HGPS}. With these improvements, a firm detection of significant extended emission in the source region is possible. The morphological model which best describes the emission is evaluated, testing both a disk-like model and a Gaussian model, each with and without the possibility of an elongation. This study finds that an elongated Gaussian model centred at $\text{R.A.}=(273.61 \pm 0.06)\degr, \text{Dec.}=(-17.39 \pm 0.07)\degr$ with a semi-major axis of $\sigma=(0.72 \pm 0.08)\degr$ describes the emission best. The gaussian model is preferred over the disk-like model with $\Delta TS = 59.91$, while the elongated Gaussian model is preferred by $3.4\,\sigma$ compared to the symmetric Gaussian model. The increased size compared to the first report of possible extended emission can be attributed to the difference in background estimation. A power-law model was used to describe the emission. This additional extended emission, hereafter referred to as HESS~J1813$-$178B, was detected with a significance of $13\,\sigma$. The best-fit spectral and spatial parameters of both models are shown in Table \ref{tab-hess:fitparamters}.  
\begin{table*}
\caption{Best-fit parameters obtained for the likelihood minimisation of H.E.S.S. data, following the spatial and spectral description introduced in section \ref{eqns_fitmodel}.} 
\label{tab-hess:fitparamters} 
\centering                          
\begin{tabular}{c | c c }       
\hline\hline  
\noalign{\smallskip}
 & HESS~J1813$-$178A & HESS~J1813$-$178B \\    
\noalign{\smallskip}
\hline 
\noalign{\smallskip}
   $\Gamma$  & $2.17 \pm 0.05_\text{stat} \pm 0.03_\text{sys}$ & $2.36 \pm 0.09_\text{stat} \pm 0.05_\text{sys}$  \\[0.1cm]
$\text{N}_0 \text{ at } 1\,\text{TeV}\,\,[10^{-12} \text{cm}^{-2}\,\text{s}^{-1}\,\text{TeV}^{-1}]$ & {$3.16 \pm 0.18_\text{stat} \pm 0.24_\text{sys}$} & {$9.89 \pm 1.40_\text{stat} \pm 1.23_\text{sys}$}   \\[0.1cm]
   R.A. $[^\circ]$ & $273.400 \pm 0.004_\text{stat} \pm 0.001_\text{sys}$ & $273.61 \pm 0.06_\text{stat} \pm 0.07_\text{sys}$  \\[0.1cm]
   Dec. $[^\circ]$ & $-17.831 \pm 0.004_\text{stat} \pm 0.001_\text{sys}$ & $-17.39 \pm 0.07_\text{stat} \pm 0.08_\text{sys}$  \\[0.1cm]
   $\sigma_M$ $[^\circ]$ & $0.056 \pm 0.003_\text{stat} \pm 0.001_\text{sys}$ & $0.72 \pm 0.08_\text{stat} \pm 0.09_\text{sys}$  \\[0.1cm]
   $e$ & - & $0.80 \pm 0.06_\text{stat} \pm 0.04_\text{sys}$  \\[0.1cm]
   $\Delta \varphi$ $[^\circ]$ & - & $40 \pm 7_\text{stat} \pm 6_\text{sys}$  \\[0.1cm]
\noalign{\smallskip}
\hline   
\hline 
\end{tabular}
\end{table*}
The best-fit position and extension along the major axis of components A and B are shown in Fig. \ref{fig-hess:morphology} by the black lines. 
The morphology of HESS~J1813$-$178B is in good spatial agreement with the extended emission  3HWC~J1813$-$174 observed by HAWC, as well as 1LHAASO~J1814$-$1719u*, extended emission observed by the WCDA of LHAASO (see figure \ref{fig-appendix:WDC_morphology}). 
\begin{figure}
\includegraphics[width=9cm]{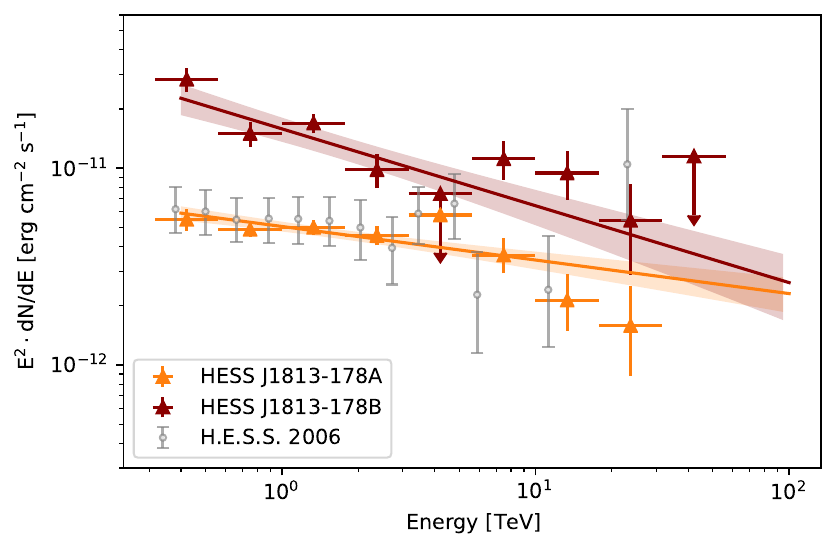}
\caption{Spectra and Spectral energy distribution (SED) of the \hess $\gamma$-ray emission from HESS~J1813$-$178A and the extended emission from HESS~J1813$-$178B. The SED derived in \citet{HGPS} are shown as filled triangles, and follow the emission from HESS~J1813$-$178A closely.}
\label{fig-hess:spectrum}
\end{figure}
The spectra of both components are depicted in Fig. \ref{fig-hess:spectrum}. Additionally, the morphological structure is tested for energy-depencence, the results of this study are presented in section \ref{enedep}.


\subsection{Fermi-\emph{LAT} results}
 
\begin{figure*}
\begin{minipage}[b]{.49\textwidth}
\includegraphics[width=\textwidth]{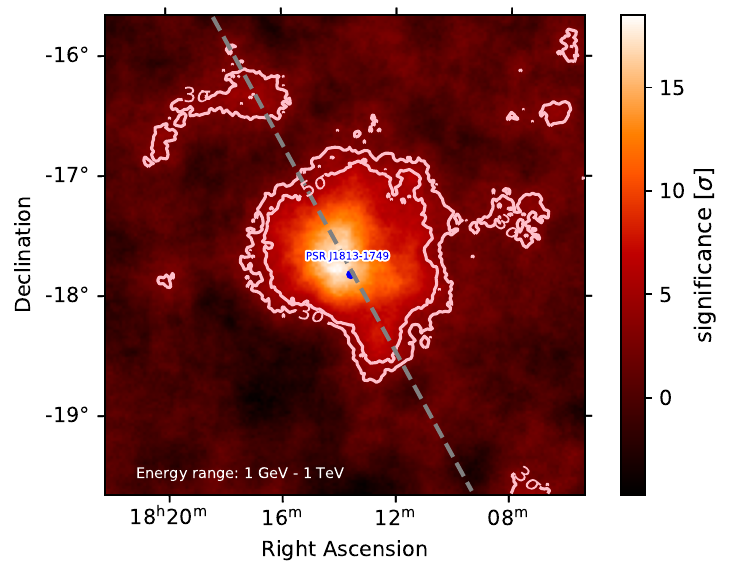}
\end{minipage}\qquad
\begin{minipage}[b]{.49\textwidth}
\includegraphics[width=\textwidth]{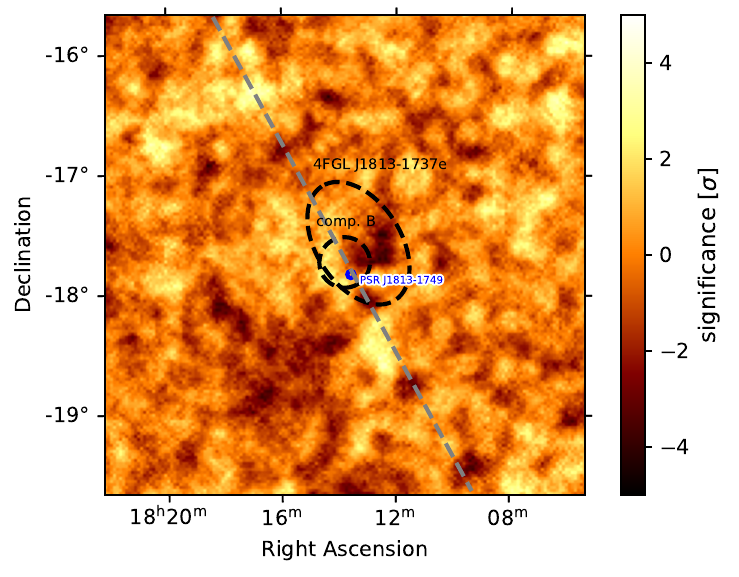}
\end{minipage}
\caption{Significance maps of the \fermilat data around the position of $4\text{FGL}\,\text{J}1813$-$1737$e in the energy range from $1\,$GeV to $1\,$TeV, with a correlation radius of  $0.1^\circ$. Left: The position of the Galactic plane is indicated by the dashed line, the pulsar position is indicated in blue, and $3\,\sigma$ and $5\,\sigma$ contours are depicted. Right: Significance map of the region after subtracting the emission using an elongated Gaussian model and a symmetric Gaussian model. The $1\,\sigma$ Gaussian extent of the models is depicted by the black dashed lines. }
\label{fig-fermi:morphology}%
\end{figure*}
\begin{figure}
\includegraphics[width=9cm]{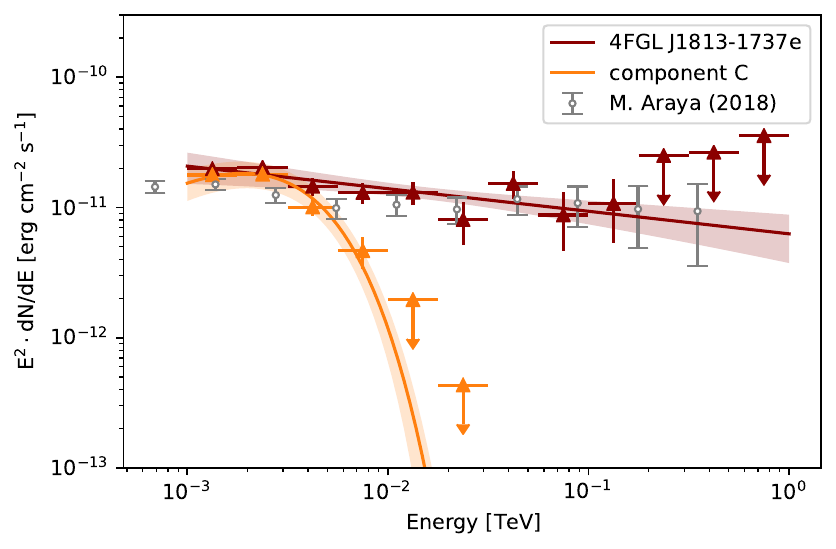}
\caption{SED and best-fit spectral models of the two-component description derived from the analysis of the data obtained by the \fermilat satellite, and SED derived in \citet{fermi_araya}.}
\label{fig-fermi:spectrum}
\end{figure}

A study using nine years of data revealed extended emission, later referred to as \fermi, in the region around \psr \citep{fermi_araya}. They reported significant, symmetric, $\gamma$-ray emission with an extension of $(0.60 \pm 0.06)^\circ$ and a power-law spectral index of $2.14 \pm 0.04$. This study reanalyses the region using an increased exposure of 12 years. 

In Fig. \ref{fig-fermi:morphology}, a significance map of the emission in the region around \fermi is shown. After the verification of the data export from \texttt{fermipy} to \texttt{gammapy} (Appendix \ref{export}), the morphological and spectral properties of the emission are estimated. The emission in the ROI is best described by an elongated Gaussian model centred at $\text{R.A.}=(273.33 \pm 0.05)\,^\circ$ and $\text{Dec.}=(-17.57 \pm 0.05)\,^\circ$, with $\sigma = (0.56 \pm 0.07)\,^\circ$. This position and size are similar to the source position of HESS~J1813$-$178B observed with H.E.S.S. (see Appendix, Fig. \ref{fig-appendix:combine_model}), indicating that the emission from \fermi and HESS~J1813$-$178B could be connected. The best-fit parameters are shown in Table \ref{fig-appendix:second_comp_tab}. 
The significance of the fitted model is $35\,\sigma$. After accounting for the emission from this component, residual emission spatially overlapping with the pulsar is observed. Therefore a second component with a symmetric Gaussian model was added to the data, to account for this residual emission. Because of the spectral shape below $10\,$GeV, a simple power-law spectral model cannot reflect the observed emission well. Therefore a power-law with exponential cut-off was chosen as a spectral model. Minimisation yields a compact component, positionally coincident with \psr. The residual significance map after the application of both models is shown on the right-hand side of Fig. \ref{fig-fermi:morphology}. Because the shape of the SED does not suggest a possible connection to HESS~J1813$-$178A, this compact component will be referred to as component C. The best-fit parameters for \fermi and component C are given in Table \ref{fig-appendix:second_comp_tab}. 
\begin{table*}
\caption{The best-fit source parameters obtained for the \fermilat data for a model consisting of an elongated Gaussian model with a power-law and a symmetric Gaussian model with an exponential cut-off power-law as a spectral model. These components are described by the equations in section \ref{eqns_fitmodel}} 
\label{fig-appendix:second_comp_tab} 
\centering                          
\begin{tabular}{c | c c }       
\hline\hline  
\noalign{\smallskip}
 & component C & \fermi \\    
\noalign{\smallskip}
\hline 
\noalign{\smallskip}
   $\Gamma$  & $ 0.95 \pm 0.22_\text{stat}$ & $2.17 \pm 0.09_\text{stat}$  \\[0.1cm]
   $\lambda$ [$\text{TeV}^{-1}$] & $ 551.5 \pm 84.4_\text{stat}$ & -  \\[0.1cm]
   $\text{N}_0  \text{ at } 2\,\text{GeV}$ $[10^{-12} \, \text{cm}^{-2}\,\text{s}^{-1}\,\text{TeV}^{-1}]$& {$8.4 \pm 2.5_\text{stat} \pm 1.83_\text{sys}$} & {$2.7 \pm 0.6_\text{stat} \pm 0.40_\text{sys}$}  \\[0.1cm]
   R.A. $[^\circ]$ & $273.452 \pm 0.004_\text{stat}$ & $273.33 \pm 0.05_\text{stat}$  \\[0.1cm]
   Dec. $[^\circ]$ & $-17.73 \pm 0.03_\text{stat}$ & $-17.57 \pm 0.05_\text{stat}$  \\[0.1cm]
   $\sigma_M$ $[^\circ]$ & $0.21 \pm 0.03_\text{stat}$ & $0.56 \pm 0.07_\text{stat}$  \\[0.1cm]
   $e$ & - & $0.78 \pm 0.07_\text{stat}$  \\[0.1cm]
   $\Delta \varphi$ $[^\circ]$ & - & $33 \pm 8_\text{stat}$  \\[0.1cm]
\noalign{\smallskip}
\hline   
\hline 
\end{tabular}
\end{table*}
The cut-off of this model was estimated to be $1.81\,$GeV. Component C improved the likelihood of the model with a significance of $3.8\,\sigma$. To detect this second component with certainty and probe the nature of this emission, events with energies below $1\,$GeV would need to be included. The SED are given in Fig. \ref{fig-fermi:spectrum}. Differences between the SED derived in this analysis and the analysis performed in \citet{fermi_araya} can be attributed to the changes from the 3FGL to the 4FGL catalogue, the usage of newly computed IRFs from Pass 8 version 3, and four years of additional data used in this analysis. 

In order to estimate the systematic uncertainties on the best-fit flux normalisation factor, two separate factors are considered. In order to estimate the error introduced by the instrument itself, the effective area has been scaled up and down by 3\%, and the analysis has been repeated. This procedure is explained in detail in \citet{fermi_sys}. The errors estimated from this method are $0.36\cdot 10^{-12} \, \text{cm}^{-2}\,\text{s}^{-1}\,\text{TeV}^{-1}$ for \fermi and $1.79\cdot 10^{-12} \, \text{cm}^{-2}\,\text{s}^{-1}\,\text{TeV}^{-1}$ for component C.
\\

In order to estimate the error introduced by the background estimation, the residual flux inside a region corresponding to the source region, at a empty position in the ROI, has been calculated. The systematic uncertainty on the number of measured counts can then be derived by dividing the total amount of residual counts by the total amount of background in the shifted region and then multiplying by the total amount of background counts in the original source region. The number of counts per bin is then scaled up and down for the whole ROI and the likelihood minimization is repeated. This process has been repeated for ten different regions. This estimation yields an systematic uncertainty of $0.18\cdot 10^{-12} \, \text{cm}^{-2}\,\text{s}^{-1}\,\text{TeV}^{-1}$ for \fermi and $0.36\cdot 10^{-12} \, \text{cm}^{-2}\,\text{s}^{-1}\,\text{TeV}^{-1}$ for component C. Both errors where then added in quadrature. 

\subsection{Energy-dependent morphology}\label{enedep}
The spatial association of the $\gamma$-ray emission with \psr, and an X-ray PWN detected by the INTEGRAL satellite, XMM-Newton and Chandra \citep{integral, xmm-newton, chandra} (Fig. \ref{fig-appendix:multiwavelenght}), suggest that a leptonic origin of the emission is plausible. Previous analyses of evolved PWNe suggest an energy-dependent morphology caused by electron diffusion and cooling \citep{J1825, enedep2, enedep1}.  To test for energy dependence of the extended emission, the data observed with H.E.S.S. was divided into three energy ranges (H1 - H3, specified in table \ref{tab:extension}) and the source models for HESS~J1813$-$178A and HESS~J1813$-$178B were fitted in each range. The \fermilat data was divided into six energy ranges (F1 - F6, also specified in table \ref{tab:extension}) and the source model for \fermi was refitted in each range. The spacing of these energy ranges was chosen to account for the small amount of statistics in the \hess data and the high-energy range in the \fermilat data. While F1 - F6 span 3 energy bins each, F6 spans 11 energy bins. The \hess data was divided such that H1 and H2 contain five energy bins of the dataset, while H3 spans 10 energy bins. 

\begin{figure}
\includegraphics[width=9cm]{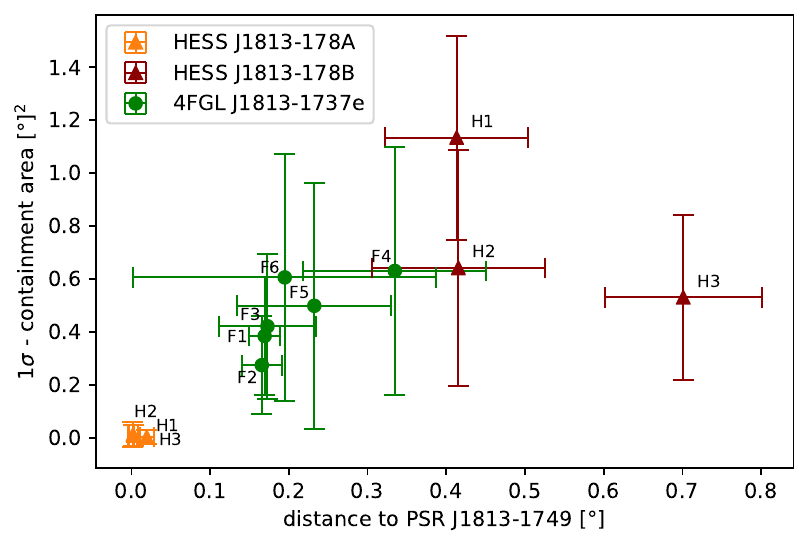}
\caption{The best-fit angular offset and area of the ellipse of HESS~J1813$-$178A and HESS~J1813$-$178B with $1\,\sigma$ uncertainties, as well as \fermi data in independent energy bins (see Table \ref{tab:extension}) H1 - H3 and F1 - F6.}
\label{fig-joint:energy-dep}
\end{figure}

Table \ref{tab:extension} shows the extension along the semi-major axis, as well as the eccentricity of the ellipse describing the emission in each energy bin and the distance of the centre of the spatial model to the pulsar. In Fig. \ref{fig-joint:energy-dep}, the dependence of the $1\,\sigma$ containment area of the ellipse and the offset to the centre of the fitted model from the pulsar position as a function of the energy is depicted. The area of the ellipse is defined as $A = \pi \sigma_M^2 \sqrt{(1 - e^2)}$. The binning was chosen based on the available statistics for each dataset.

The enclosed area is compatible within errors in the overlapping energy ranges of the \fermilat and H.E.S.S. data. There is no significant indication for energy-dependence of the best-fit containment area of HESS~J1813$-$178A and HESS~J1813$-$178B. This study does however reveal a dependence of the distance between the pulsar and the centre of the emission, which increases with increasing energy.  
As the morphology of HESS\,J1813$-$178A remains fixed throughout all energy bands, the increased offset of the best-fit position towards higher energies may indicate that the particle transport occurs preferentially towards a single direction before spreading out more isotropically as the particles lose energy and cool. Alternatively, this could indicate that a second faint source is present in the region. 

\begin{table*}
\caption{The best-fit parameters for the morphology of HESS~J1813$-$178B in different energy bands, for both \fermilat (F1 - F6) and \hess datasets (H1 large - H3 large) An elongated Gaussian model is assumed in all cases and the $1\,\sigma$ statistical uncertainties are given. Additionally, the best-fit values for the symmetrical Gaussian used to account for the emission from HESS~J1813$-$178A.}
\label{tab:extension} 
\centering                          
\begin{tabular}{c c | c c c c }       
\hline\hline  
\noalign{\smallskip}
& Energy [GeV] & RA $[^\circ]$ & Dec $[^\circ]$ & $\sigma_M$ [$^\circ$] & $e$  \\
\noalign{\smallskip}
\hline 
\noalign{\smallskip}
   F1 & $1.0\,-\,2.0$ & $273.46  \pm  0.02$ & $-17.67  \pm  0.02$ & $0.39 \pm 0.04$ & $0.81 \pm 0.05$ \\
   F2 & $2.0\,-\,4.0$ & $273.48  \pm  0.03$ & $-17.69  \pm  0.03$ & $0.42 \pm 0.04$ & $0.85 \pm 0.04$ \\
   F3 & $4.0\,-\,7.5$ & $273.38  \pm  0.04$ & $-17.66  \pm  0.06$ & $0.39 \pm 0.08$ & $0.87 \pm 0.07$ \\
   F4 & $7.5\,-\,18$  & $273.26  \pm  0.09$ & $-17.52  \pm  0.12$ & $0.53 \pm 0.11$ & $0.75 \pm 0.19$ \\
   F5 & $18\,-\,58$ & $273.17  \pm  0.10$ & $-17.77  \pm  0.08$  & $0.57 \pm 0.16$ &  $0.77 \pm 0.21$ \\
   F6 & $58\,-\,1.0 \times 10^{3}$ & $273.22  \pm  0.20$ & $-17.75  \pm  0.13$ & $0.71 \pm 0.19$ & $0.92 \pm 0.08$ \\
    \noalign{\smallskip}
   \cdashline{1-6}
   \noalign{\smallskip}
   H1 small & $(0.4\,-\,1.3) \times 10^{3}$ & $273.393  \pm  0.005$ & $-17.832  \pm  0.005$  & $0.054 \pm 0.004$ & $--$ \\
   H2 small & $(1.3\,-\,5.7) \times 10^{3}$ & $273.397  \pm  0.006$ & $-17.834  \pm  0.007$ & $0.064 \pm 0.005$ & $--$ \\
   H3 small & $(5.7\,-\,100) \times 10^{3}$ & $273.408  \pm  0.009$ & $-17.817  \pm  0.009$ & $0.035 \pm 0.006$ & $--$ \\
    \noalign{\smallskip}
   \cdashline{1-6}
   \noalign{\smallskip}
   H1 large & $(0.4\,-\,1.3) \times 10^{3}$ & $273.61  \pm  0.08$ & $-17.48  \pm  0.09$ & $0.87 \pm 0.12$ & $0.88 \pm 0.04$ \\
   H2 large & $(1.3\,-\,5.7) \times 10^{3}$ & $273.41  \pm  0.09$ & $-17.42  \pm  0.11$  & $0.48 \pm 0.09$ & $0.48 \pm 0.38$ \\
   H3 large & $(5.7\,-\,100) \times 10^{3}$ & $273.65  \pm  0.11$ & $-17.18  \pm  0.10$ & $0.41 \pm 0.09$ & $0.00 \pm 0.02$ \\
\noalign{\smallskip}
\hline     
\hline 
\end{tabular}
\end{table*}

\subsection{Joint Model Fit}
\label{joint-analysis}
A consistent description of the ROI through the GeV-TeV range can be reached by fitting the \fermilat and \hess data jointly. Following the results derived in the analysis of the respective datasets, three source models were used to describe the data across five decades of energy. First a symmetric Gaussian model, as well as an elongated Gaussian model was added to both datasets. These models will be referred to as component A and B respectively. Additionally, a symmetric Gaussian model with an exponential cut-off power-law spectral model was added only to the \fermilat data. This model component, referred to as component C in the analysis of the \fermilat data. The likelihood minimisation was then performed on both datasets at the same time. The best-fit parameters are shown in Table \ref{tab-joint:parameters}.
\begin{table*}[ht]
\caption{The best-fit parameters derived in the joint analysis of the H.E.S.S. and \fermilat data.} 
\label{tab-joint:parameters} 
\centering                          
\begin{tabular}{c | c c c }       
\hline\hline  
\noalign{\smallskip}
 & comp. A & comp. B & comp. C  \\    
\noalign{\smallskip}
\hline 
\noalign{\smallskip}
   $\Gamma$  & $2.05 \pm 0.03$ & $2.17 \pm 0.03$ & $-0.12 \pm 0.02$  \\
   $\beta$ & $0.0620 \pm 0.003$ & $0.0437 \pm 0.008$ & -  \\ 
    $\alpha$ & - & - & $0.66 \pm 0.01$ \\
   $\lambda$ [$10^{3} \cdot \text{TeV}^{-1}$] & -  & -  & $ 3.93 \pm 0.20$ \\
   $\text{N}_0  \text{ at } 1\,\text{TeV}$ $[10^{-12} \, \text{cm}^{-2}\,\text{s}^{-1}\,\text{TeV}^{-1}]$ & {$3.16 \pm 0.14$} & {$6.47 \pm 0.43$} & {$9.89 \pm 1.64$}   \\
   \noalign{\smallskip}
   \cdashline{1-4}
   \noalign{\smallskip}
   RA $[^\circ]$ & $273.400 \pm 0.003$ & $273.39 \pm 0.03$  & $273.48 \pm 0.02$ \\
   Dec $[^\circ]$ & $-17.832 \pm 0.003$ & $-17.50 \pm 0.04$ & $-17.67 \pm 0.02$ \\
   $\sigma_M$ $[^\circ]$ & $0.056 \pm 0.003$ & $0.54 \pm 0.03$ & $0.29 \pm 0.02$  \\
   $e$ & - & $0.73 \pm 0.04$ & -  \\
   $\Delta \varphi$ $[^\circ]$ & - & $33 \pm 7$ & -  \\
\noalign{\smallskip}
\hline   
\hline 
\end{tabular}
\end{table*}

\begin{figure}
\includegraphics[width=9cm]{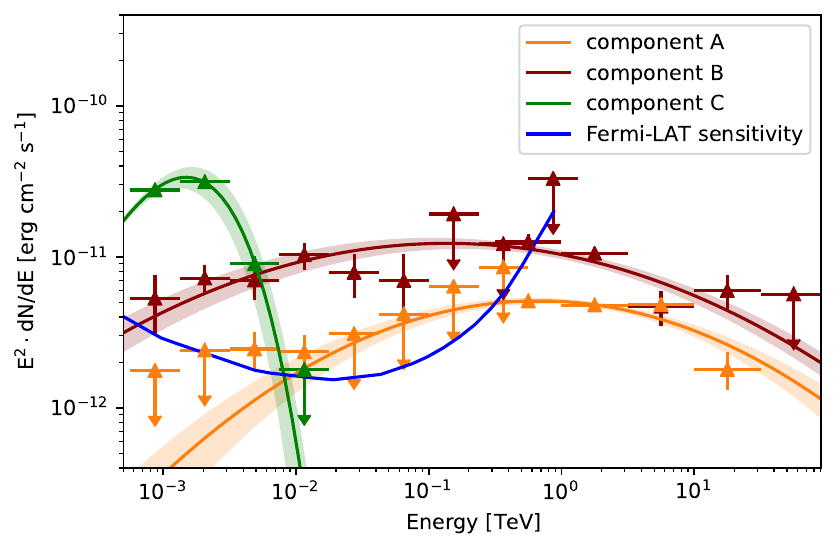}
\caption{SED from a joint-analysis of the combined H.E.S.S. and \fermilat data. 
The 12 year sensitivity of the LAT is indicated by the blue line.}
\label{fig-joint:spectrum}
\end{figure}

Figure \ref{fig-joint:spectrum} depicts the spectra and SED of the best-fit models, together with the sensitivity of the LAT for 12 years of exposure in blue. The curve shows the broadband sensitivity for sources located in the Galactic plane \citep{sensitivity}. While the best-fit parameters of the models where estimated using the data from both datasets at the same time, the flux points were computed in the respective datasets. This results in a energy range where both the \hess and \fermilat data show significant flux or upper limits. 
If the spectrum of component A does not change drastically at the lower energies, measuring it in the LAT data would not be possible as components A and C are positionally coincident and the detector sensitivity is insufficient. Together these three different source models describe the emission of the ROI well. The advantage of this joint model becomes evident in the GeV energy regime. The spatial separation between \fermi and the compact emission component detected in the \fermilat data is complicated, but using the assumption that \fermi and HESS~J1813$-$178B have the same origin, the influence of the second component on the model parameters can be reduced. This becomes evident when comparing the flux around $1\,$GeV, which has been reduced by a factor two in the joint model. 

\section{Spectral Modeling using GAMERA}
Two scenarios were investigated for the origin of the extended $\gamma$-ray emission in the region of \psr. In the first scenario, a magnetised wind of charged particles, primarily electrons and positrons, originating from \psr, forms a PWN. This leptonic emission scenario has been favoured in previous analyses of the X-ray and TeV emission \citep{integral, xmm-newton}. In systems older than $10\,$kyr, the electrons can escape the confines of the PWNe and diffuse through the ISM, although the age at which this occurs may vary for individual systems. The diffuse $\gamma$-ray emission produced by these electrons is typically observed around middle-aged pulsars and one can distinguish between a majority of the particles still confined in the PWN and one where most particles have escaped into the ISM \citep{pwn_population}. This evolutionary stage of a leptonic system could potentially correspond to the extended emission observed as HESS~J1813$-$178B.

Another possible origin of the extended emission are cosmic-ray nuclei, accelerated at shock fronts and which have since escaped from their accelerator. A possible candidate for such an acceleration site is the SNR $\text{G}12.82-0.02$, which is positionally coincident with the pulsar \psr and believed to be the pulsars host SNR (see Fig. \ref{fig-appendix:multiwavelenght}). After acceleration and propagation, the nuclei may interact within dense molecular clouds in the region, producing high-energy $\gamma$-rays. 

This emission scenario has been favoured in an analysis of nine years of \fermilat data, performed by \citet{fermi_araya}. He suggested the stellar cluster $\text{Cl}\,\text{J}1813$-$178$ as a possible origin, but also states that the extension of the $\gamma$-ray emission exceeds the extension of the star-forming region, and therefore an unknown component of star formation would be necessary. 

\subsection{Leptonic scenario}\label{lept}
To describe the $\gamma$-ray emission in the region using a time-dependent model of electrons originating from \psr, the GAMERA package \citep{gamera} was used. 
\begin{table}
\caption{The assumed properties of \psr following newest estimates from \citet{psr_distance1}. For the braking index and break energy canonical values are assumed.} 
\label{tab-modeling:pwn_properties} 
\centering                          
\begin{tabular}{c | c  }
\hline 
\hline 
\noalign{\smallskip}
    Parameter & Assumed Value \\
    \noalign{\smallskip}
\hline 
\noalign{\smallskip}
    Distance, d & $6.2\,$kpc \\
    spin-down power, $\dot{E}$ & $5.6\times10^{37}\,$erg/s \\
    spin period, $P$ & $44.7\cdot10^{-3}\,$s \\
    change in spin period, $\dot{P}$ & $1.27\times10^{-13}\,$s/s \\
    characteristic age $\tau_c$ & $5.585\,$kyr \\
    Braking index, $n$ & $3.0\,$ \\
    Break energy, $E_\text{b}$ & $100\,$GeV \\
\noalign{\smallskip}
\hline  
\hline 
\end{tabular}
\end{table}

Additionally to the SED derived in the joint analysis, measurements of the region in the $20 - 100\,$keV band, showing an unresolved hard X-ray source observed by the INTEGRAL satellite \citep{integral}, XMM-Newton \citep{xmm-newton} (with an extraction region of $75\arcsec$), and Chandra \citep[with an elliptical extraction region of $6\arcsec \times 8\arcsec$]{chandra} were used, as well as observation data from the Very Large Array in the $20\,$cm and $90\,$cm band, revealing a shell-like non-thermal radio source identified as SNR G12.82–0.02 \citep{VLA}. The radio data is only shown for comparison and was not included in the modeling.

The SED used for this model were not derived using the same spatial extent. While one possible solution for this discrepancy would be to scale the observed flux for the different observation regions, this would require the assumption that a linear scaling can be applied. Additionally, X-ray emission generally traces higher energy electrons, and hence younger electrons, than the TeV emission. To account for these differences, three electron ‘generations’ were defined, following the procedure in \citet{J1809}. The first electron generation, the relic electrons, were injected since the birth of the pulsar and are now detected as extended high-energy $\gamma$-ray emission. The second electron generation, the middle-aged electrons, were injected since $t_e = t_\text{true} \cdot t_\text{PWN}$, with $t_\text{true}$ the age of the system, and are now detected as PWN around \psr. The third electron generation, the young electrons injected since  $t_e = t_\text{true} \cdot t_\text{X-ray}$ are responsible for the X-ray synchrotron emission. 
\begin{figure*}
\begin{minipage}[b]{.49\textwidth}
\includegraphics[width=\textwidth]{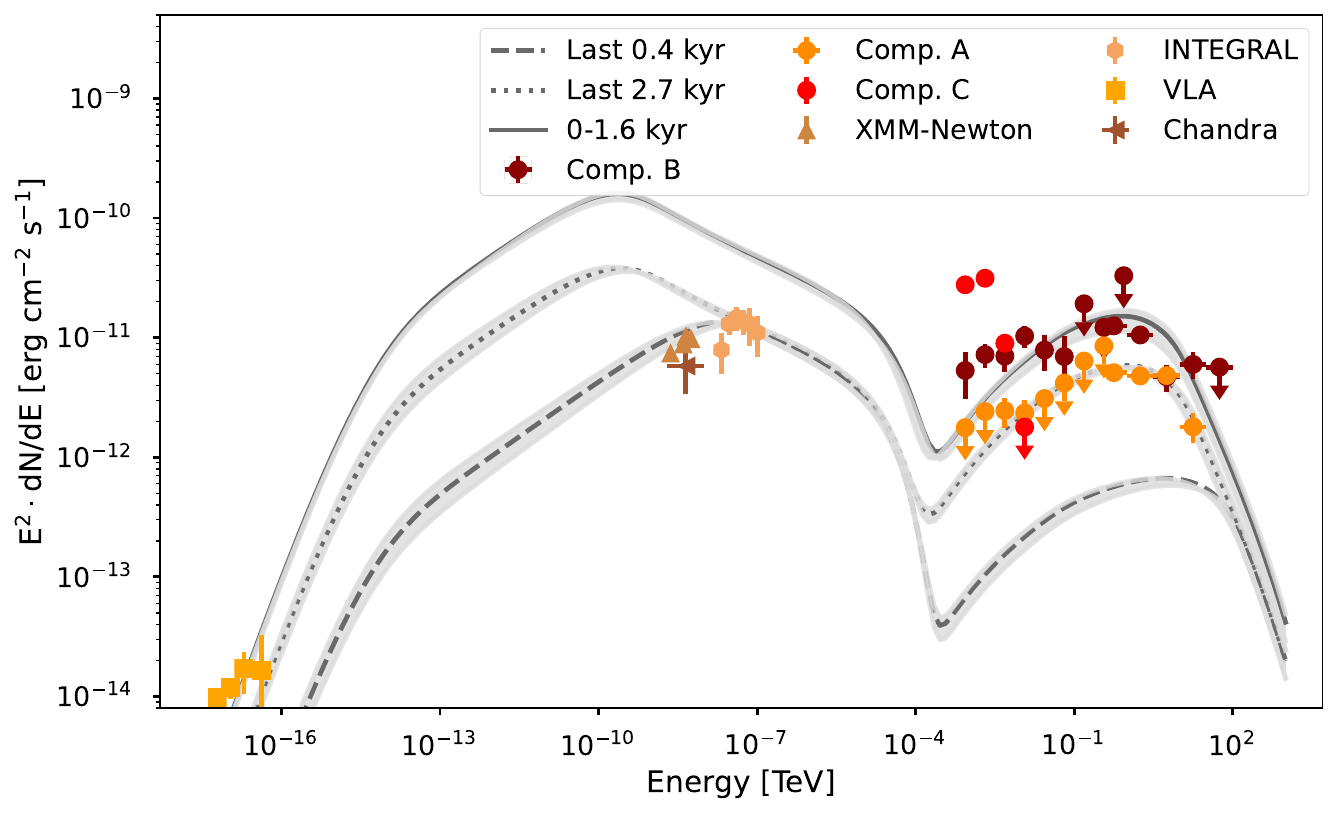}
\end{minipage}\qquad
\begin{minipage}[b]{.49\textwidth}
\includegraphics[width=\textwidth]{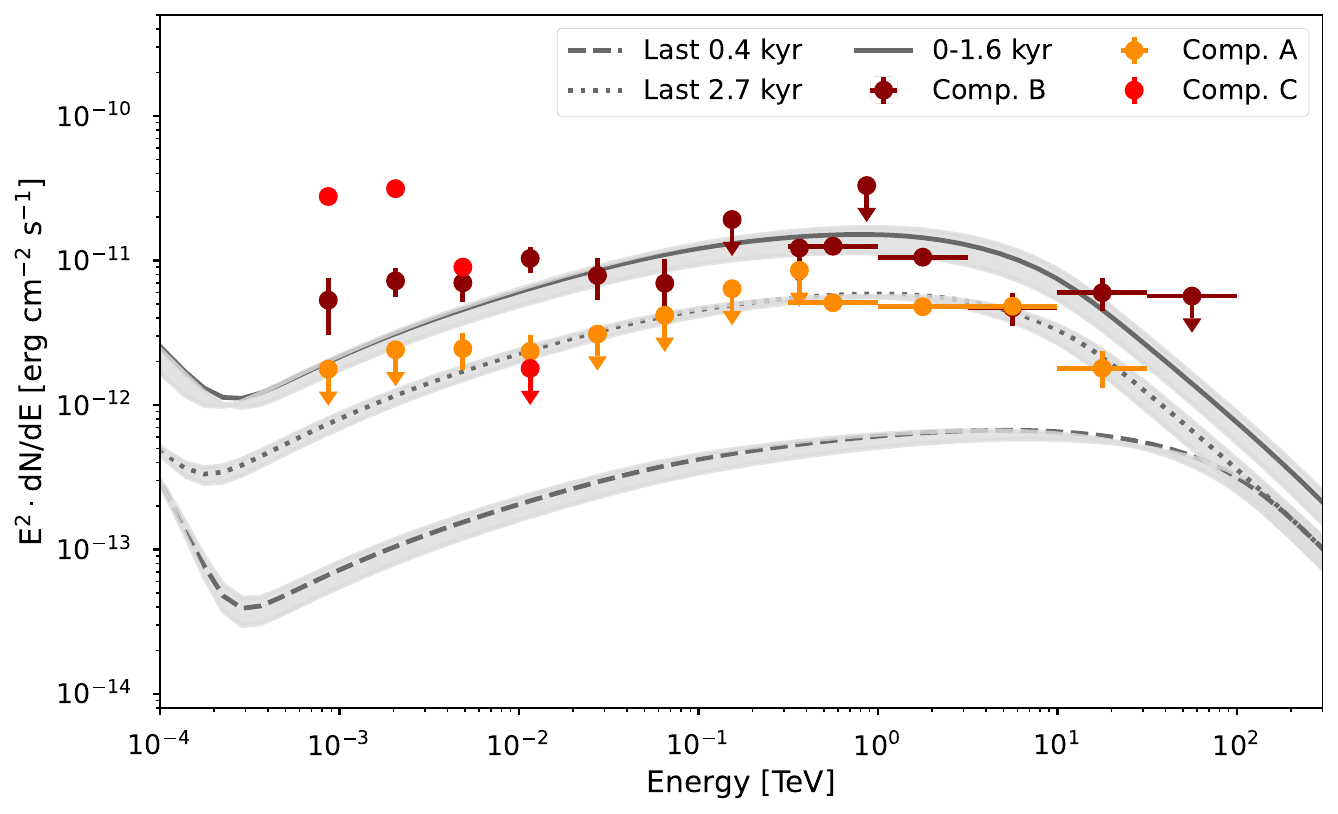}
\end{minipage}
\caption{SED derived in this analysis, as well as X-ray SED by XMM-Newton \citep{xmm-newton} Chandra \citep{chandra} and INTEGRAL \citep{integral} and SED derived in the analysis of VLA data \citep{VLA}, compared to the SED curves expected from the leptonic model obtained with GAMERA. Left: Full energy range. Right: Comparison between the $\gamma$-ray flux estimated from the leptonic model from $0-1.6\,$kyrs and the last $2.5\,$kyrs and the emission observed from HESS~J1813$-$178B and HESS~J1813$-$178A respectively. The shaded grey error bands indicate the possible parameter space, given in Table \ref{tab-modeling:fit-parameters}.}
\label{fig-modeling:model}
\end{figure*}
For this model, an electron injection spectrum following a power-law with a spectral index of $\alpha$, a break energy of $E_b$ and a cutoff $E_{\rm cut}$:
\begin{linenomath}
\begin{ceqn}
\begin{equation}
N(E_{\rm e}) = \left( 1 + \frac{E_e}{E_b}\right) ^\alpha \cdot \exp\left( \frac{-E_{\rm e}}{E_{\rm cut}}\right)\,,
\end{equation}
\end{ceqn}
\end{linenomath}
where $E_e$ the electron energy, was used. 

The distance to the pulsar, as well as pulsar-specific parameters, such as spin-down period $\text{P}$, change in spin period $\dot{\text{P}}$, and spin-down power $\dot{\text{E}}$ were taken from \citet{psr_distance1}. All assumed properties of the pulsar used for this model are given in Table \ref{tab-modeling:pwn_properties}.

The injection index $\alpha$, the birth period $\text{P}_0$, and the conversion fraction of the spin-down luminosity to electrons $\vartheta$, as well as the magnetic field present at the time of the observation of the $\gamma$-ray emission $\text{B}_\text{now}$, and the age of the electron populations were free parameters of the model.  The spin-down power of the pulsar was assumed to evolve in time with:
\begin{linenomath}
\begin{ceqn}
\begin{equation}
\dot{\text{E}}=\dot{\text{E}}_0\left(1+\frac{t}{\tau_0}\right)^{-2} \,,
\end{equation}
\end{ceqn}
\end{linenomath}
with $\tau_0 = \text{P}_0/(2\,\dot{\text{P}_0})$, the spin-down time at the birth of the pulsar and $\dot{\text{E}}_0$ the initial spin-down energy loss rate at birth of the pulsar. Similarly the pulsar period was assumed to evolve as $\text{P(t)}=\text{P}_0\left(1+\frac{t}{\tau_0}\right)^{0.5}$, with $\text{P}_0$ the birth period of the pulsar and the magnetic field as \(\text{B(t)}=\text{B}_0\left(1+\left(\frac{t}{\tau_0}\right)^{0.5}\right)^{-1}\) \citep{modeling2, model3} with $\text{B}_0$ the magnetic field at birth of the pulsar. The spin down luminosity is defined as:
\begin{linenomath}
\begin{ceqn}
\begin{equation}
\text{L}=\text{L}_0 \cdot \vartheta \cdot \left(1 + \frac{t}{t_0}\right)^{-2}
\end{equation}
\end{ceqn}
\end{linenomath}
with the inital spin down luminosity $L_0 = \dot{\text{E}}\cdot (1 + (t_\text{true}/t_0))^2$ and $t_0 = \frac{P_0^2}{2 P \dot{P}}$.

Since the time at which the escape of the particles into the ISM first occurs is unknown, this modeling approach assumes that the particles have been able to escape instantly and therefore the relic electrons have been injected since the birth of the pulsar. An additional caveat of this model is, that while this model is time-dependent and cooling losses of electrons were taken into account, no spatial dependence or spatial evolution was added. The fit of the different electron generations was performed only on the SED. 

The free parameters of the model were fitted to the observation data from the X-ray satellites, as well as the SEDs derived in the joint-analysis from components A and B. For optimisation, the package \emph{EMCEE} \citep{emcee}, which applies a Markov chain Monte Carlo (MCMC) method, was used.

The optimised model is shown in Fig. \ref{fig-modeling:model}, 
while the parameter space that yields the highest numerical probability ($16 - 84$\% range of the probability distribution) for the MCMC method is listed in Table \ref{tab-modeling:fit-parameters}. This morphology independent model can account well for the observed emission. 

The observed X-ray data in the vicinity of \psr can be well explained by young electrons that were produced only in the last $0.4\,$kyr, the electrons of the third generation. The electrons produced in the last $2.7\,$kyr, the second generation, were observed by H.E.S.S. as compact emission around the pulsar. 
\begin{table}
\caption{Validity range of the parameters used in the evolution of the leptonic model. The parameter combination presented here should be interpreted as a range of possible combinations, not best-fit values.
} 
\label{tab-modeling:fit-parameters} 
\centering                          
\begin{tabular}{c | c  }  
\hline 
\hline 
\noalign{\smallskip}
Parameter & Validity range \\
\noalign{\smallskip}
\hline 
\noalign{\smallskip}
    $B_\text{now}$ & [$11.36$ $-$ $13.18$]$\,\mu$G \\
    $P_0$ & [$20.19$ $-$ $22.97$ ]$\,$ms \\
    $\vartheta$ & [$0.15$ $-$ $0.17$]$\,$ \\
    $\alpha$ & [$2.35$ $-$ $2.40$]$\,$ \\
    $\Delta t_\text{X-ray}$ & [$0.88$ $-$ $0.91$]$\,$ \\
    $\Delta t_\text{PWN}$ & [$0.35$ $-$ $0.39$]$\,$ \\
\noalign{\smallskip}
\hline    
\hline 
\end{tabular}
\end{table}
Taking into account the low statistics at high energies, the population of relic electrons from the first generation, released from the birth of the pulsar up to an age of $1.6\,$kyrs, can describe the extended emission from HESS~J1813$-$178B reasonably well but fails to describe the emission from component C. To describe this emission in a leptonic scenario, a second electron population and an unreasonably high particle density, only experienced by the second electron generation, would need to be introduced. 

Using the parameters of the model, the true age of the system was estimated, following \citet{modeling2}:

\begin{linenomath}
\begin{ceqn}
\begin{equation}
\label{eq:age}
    t_\text{true} = \frac{\text{P}}{(n-1)\dot{\text{P}}}\left(1- \left(\frac{\text{P}_0}{\text{P}}\right)^{n-1}\right)\,\,.
\end{equation}
\end{ceqn}
\end{linenomath}
with $n$ the braking index of the pulsar. For the validity range of parameters in this analysis, the true age of the system was estimated to be $\sim 4.7\,$kyrs. 
\begin{figure}
\includegraphics[width=9cm]{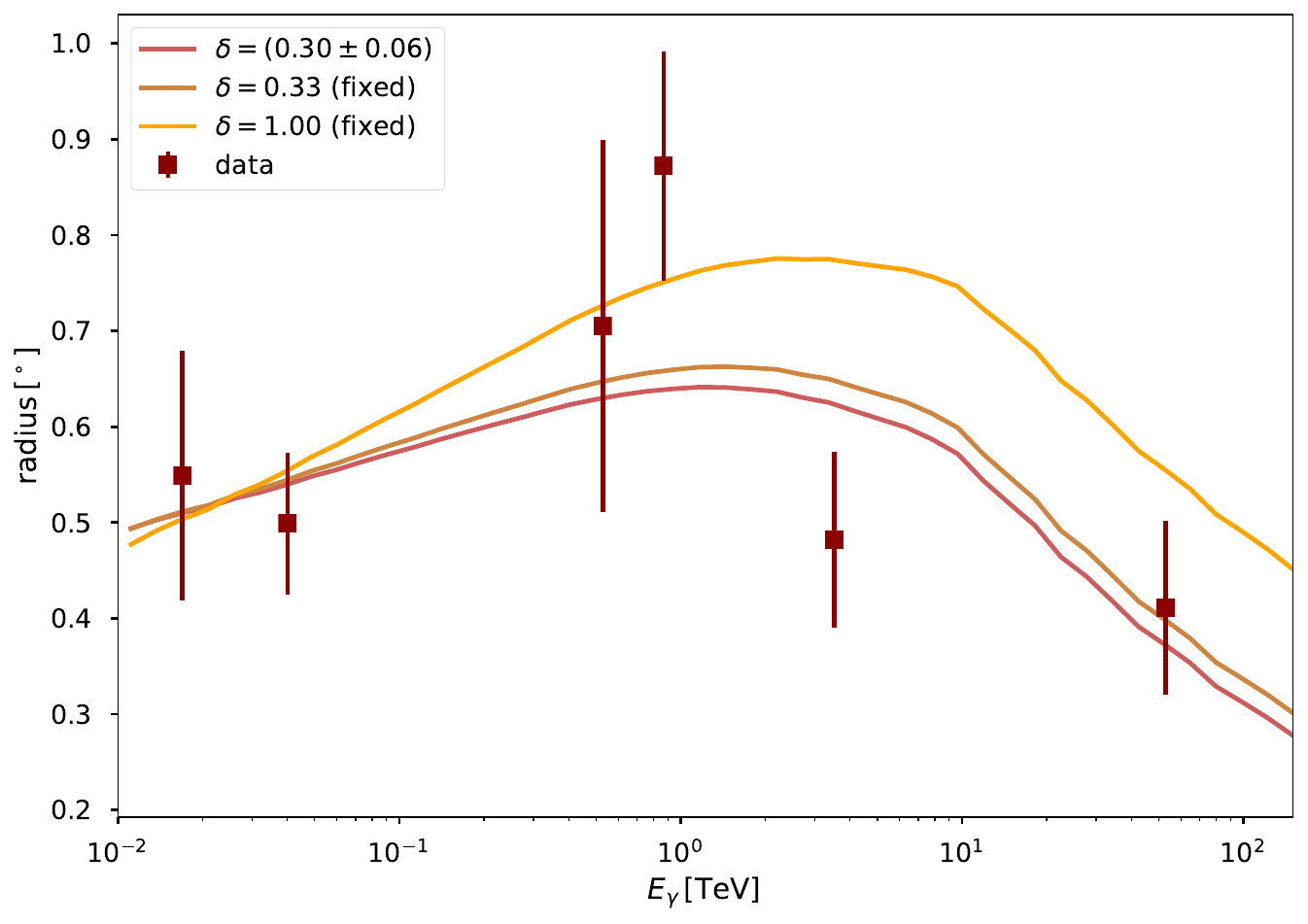}
\caption{The measured extension of HESS~J1813$-$178B in continuous energy bands, specified in table \ref{tab:extension}, is shown in red. The figure also shows the expected $1\,\sigma$ containment radius for different diffusion coefficients.
}
\label{fig-modeling:diff_coeff}
\end{figure}
With this information, together with the measurement of the extension of \fermi and HESS~J1813$-$178B in the energy bands specified in table \ref{tab:extension}, an estimation of the diffusion coefficient and diffusion index can be made by comparing the observed angular size to that expected for the relic electron population at different $\gamma$-ray energies. In order to estimate a diffusion coefficient radial symmetric diffusion was assumed. To get a estimation of the radial symmetric extend of the the emission from HESS~J1813$-$178B, a symmetric Gaussian model was fitted to the emission in the energy bands defined in Section \ref{enedep}.  This derived extension, as well as the age of the system, derived using equation \ref{eq:age}, were then used to estimate the distance which the relic electrons have diffused since they were injected, with the expected angular size calculated as $r_\text{diff} = \sqrt{2 t_\text{age} \cdot D}$, with $D$ the diffusion coefficient. The diffusion coefficient was derived using the electron energy $\text{E}_e$, the diffusion index $\delta$ and the diffusion coefficient $\text{D}_0$ at the reference energy of $1\,$TeV:

\begin{linenomath}
\begin{ceqn}
\begin{equation}
\label{eq:diff}
    \text{D} = \text{D}_0\left(\frac{\text{E}_e}{1\,\text{GeV}}\right)^\delta\,\,.
\end{equation}
\end{ceqn}
\end{linenomath}
The diffusion coefficient is estimated to $\text{D}_0 = (6.98 \pm 0.69) \times 10^{28}\,\text{cm}^2/\text{s}$ with a diffusion index $\delta = (0.30 \pm 0.06)$. The estimated radius as a function of $\gamma$-ray energy, as well as the measurement, can be seen in Fig. \ref{fig-modeling:diff_coeff}. 
The estimated diffusion index lies in the range between Kraichnan turbulence ($\delta=0.5$, \citep{kraichnan}) and Kolmogorov turbulence ($\delta=0.33$, \citep{kolmogorov}), which is the canonical diffusion index typically assumed for PWNe. The best-fit diffusion coefficient is comparable with the diffusion coefficient estimated for the ISM \citep{ISM_diff}. 

To compare the results to the population of known pulsars, the energy density of both HESS~J1813$-$178A and HESS~J1813$-$178B in the TeV energy range was estimated, following the scheme in \citet{pwn_population}. The energy density of the electrons was estimated using $\epsilon_e  = \frac{\text{E}_\text{inj}}{\text{V}}$, or using the properties of the pulsar

\begin{linenomath}
\begin{ceqn}
\begin{eqnarray}
\label{eq:density}
 \epsilon_e & = & \frac{\dot{\text{E}}\tau_c}{4\pi (\text{R}_1^2 \cdot \text{R}_2) /3}\,\,,
\end{eqnarray}
\end{ceqn}
\end{linenomath}
with $\dot{E}$ the present-day spin-down power, $\tau_c$ the present-day characteristic age, $\text{R}_1$ the extension of the semi-major axis of the ellipse and $\text{R}_2$ the size the semi-minor axis. The energy density using the total TeV $\gamma$-ray luminosity was estimated by integrating over the electron population inferred by our best-fit model. This study finds $E_\text{inj} = $3.62$ \cdot 10^{48}\,$erg. The derived electron densities are given in Table \ref{tab-modeling:densities}. The energy density calculated from the $\gamma$-ray luminosity is lower than that estimated from the properties of the pulsar. This effect was already observed by \citet{pwn_population}, a possible reason could be that only the estimation using the $\gamma$-ray luminosity takes the evolution into account, while the estimation using the current pulsar properties cannot sufficiently describe this evolution. The energy density calculated for HESS~J1813$-$178A is compatible with the energy densities estimated for other PWNe, while the energy density of HESS~J1813$-$178B is close to $0.1\,\text{eV}/\text{cm}^3$ and therefore comparable to the values estimated for pulsar halos \citep{pwn_population}.
\begin{table}
\caption{Estimated electron density following equation (\ref{eq:density}).}
\label{tab-modeling:densities} 
\centering                          
\begin{tabular}{c | c c }  
\hline 
\hline 
\noalign{\smallskip}
     $\epsilon_e\,\, [\text{eV}/\text{cm}^3]$ & HESS~J1813$-$178A & HESS~J1813$-$178B \\
     \noalign{\smallskip}
    \hline 
    \noalign{\smallskip}
    estimated from $\dot{\text{E}}$ & $65.29$ $\pm$ $4.49$ & $0.11$ $\pm$ $0.02$\\
    estimated from $\text{L}_\gamma$ & $23.95$ $\pm$ $1.64$ & $0.04$ $\pm$ $0.01$ \\
\noalign{\smallskip}
\hline    
\hline 
\end{tabular}
\end{table}

\subsection{Hadronic origin}
Due to the positional coincidence of the emission and the SNRs G012.8–00.0 and G012.7–00.0 or the star-forming region W33, there is also the possibility that the observed emission is a superposition of $\gamma$-ray emission produced by different sources that overlap along the line of sight. While there is no information available for SNR G012.7-00.0, the distance to \psr and the SNR G012.8–00.0 is estimated to be $6.2 - 12\,$kpc \citep{psr_distance1}, while the stellar cluster is expected to be located at a much smaller distance of $4.8\,$kpc \citep{stellar_cluster}. It is therefore possible that the emission observed as HESS~J1813$-$178A is a pulsar wind nebula powered by \psr, while the extended emission observed in the \fermilat and \hess data is caused by protons accelerated at the SNR shock front or the stellar cluster.

\begin{figure}
\includegraphics[width=9cm]{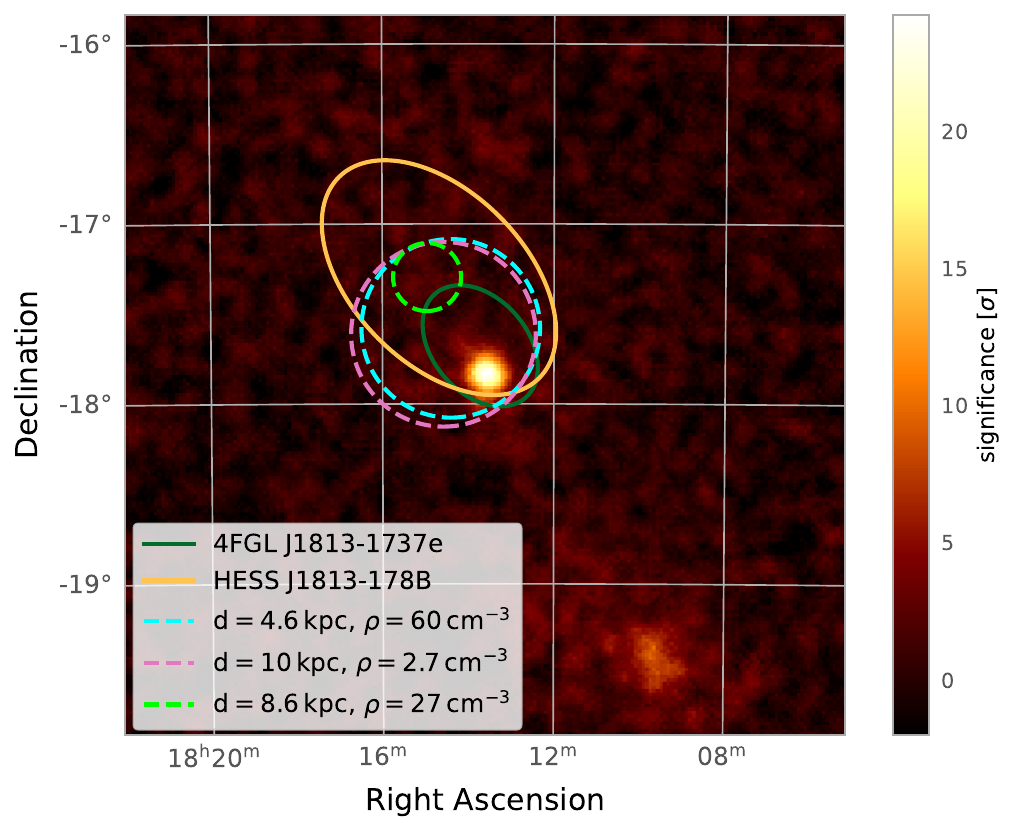}
\caption{Significance map computed from the H.E.S.S data with an correlation radius of $0.06^\circ$. The estimated position and density of molecular clouds in the region are indicated by the dashed lines. Additionally, the morphology of the best-fit model derived in the analysis of the \fermilat and \hess data is indicated. Only clouds with a distance between $4\,$kpc and $12\,$kpc and a large positional overlap with the extended emission observed in \fermilat and \hess are depicted in the counts map.}
\label{fig-modeling:clouds}
\end{figure}

In order to produce high energy $\gamma$-rays via hadronic interactions, target material is necessary. Using the molecular cloud catalogues provided in \citet{clouds2} and \citet{clouds1}, three molecular clouds, positionally coincident with the best-fit position of the emission from HESS~J1813$-$178B were identified. Their location compared to the emission observed in the analysis of the \fermilat data can be seen in Fig. \ref{fig-modeling:clouds}. 
In order to investigate the feasibility of this scenario as the origin of the emission observed from \fermi and HESS~J1813$-$178, a proton and electron population, accelerated at the shock front of the SNR were evolved over time. Following \citet{hadron_injection}, the protons were injected with: 
\begin{linenomath}
\begin{ceqn}
\begin{eqnarray}
\label{eq:spec_proton}
 f(p) & = & N_p \cdot E^{-\alpha_p} \cdot \exp{\left(\frac{-E}{E_\text{max}}\right)} \cdot \exp{\left(-\frac{E_\text{min}}{E}\right)}
\end{eqnarray}
\end{ceqn}
\end{linenomath}
where $N_p$ is the normalisation factor, $E$ is the energy of the injected particles, $E_\text{min} = m_p$, the rest mass of the proton, $E_\text{max}$ is assumed to be $1\,$PeV, and $\alpha_p$ is the spectral index of the injected protons. The injection spectrum of the electrons was defined as:
\begin{linenomath}
\begin{ceqn}
\begin{eqnarray}
\label{eq:spec_combined}
 f_{e}(p) & = & N_p \cdot k_{ep} \cdot E^{-(\alpha_p + \Delta \alpha)} \cdot \exp{\left(\frac{E}{E_\text{max}}\right)} \cdot \exp{\left(-\frac{E_\text{min}}{E}\right)}
\end{eqnarray}
\end{ceqn}
\end{linenomath}
with $\Delta \alpha$ the difference between the injection index of the electrons and protons, $k_{ep}$ the electron-to-proton ratio, and $E_\text{min} = 100m_e$, with the electron rest mass, as well as $E_\text{max} = 1\,$TeV. This particle population is evolved in an environment with the ambient magnetic field $B_\text{now}$ and the particle density $d$. We assume a particle density of $d = 60\,\text{cm}^{-3}$, which corresponds to the density of one of the molecular clouds in the region (see figure \ref{fig-modeling:clouds}). 

\begin{figure}
\includegraphics[width=9cm]{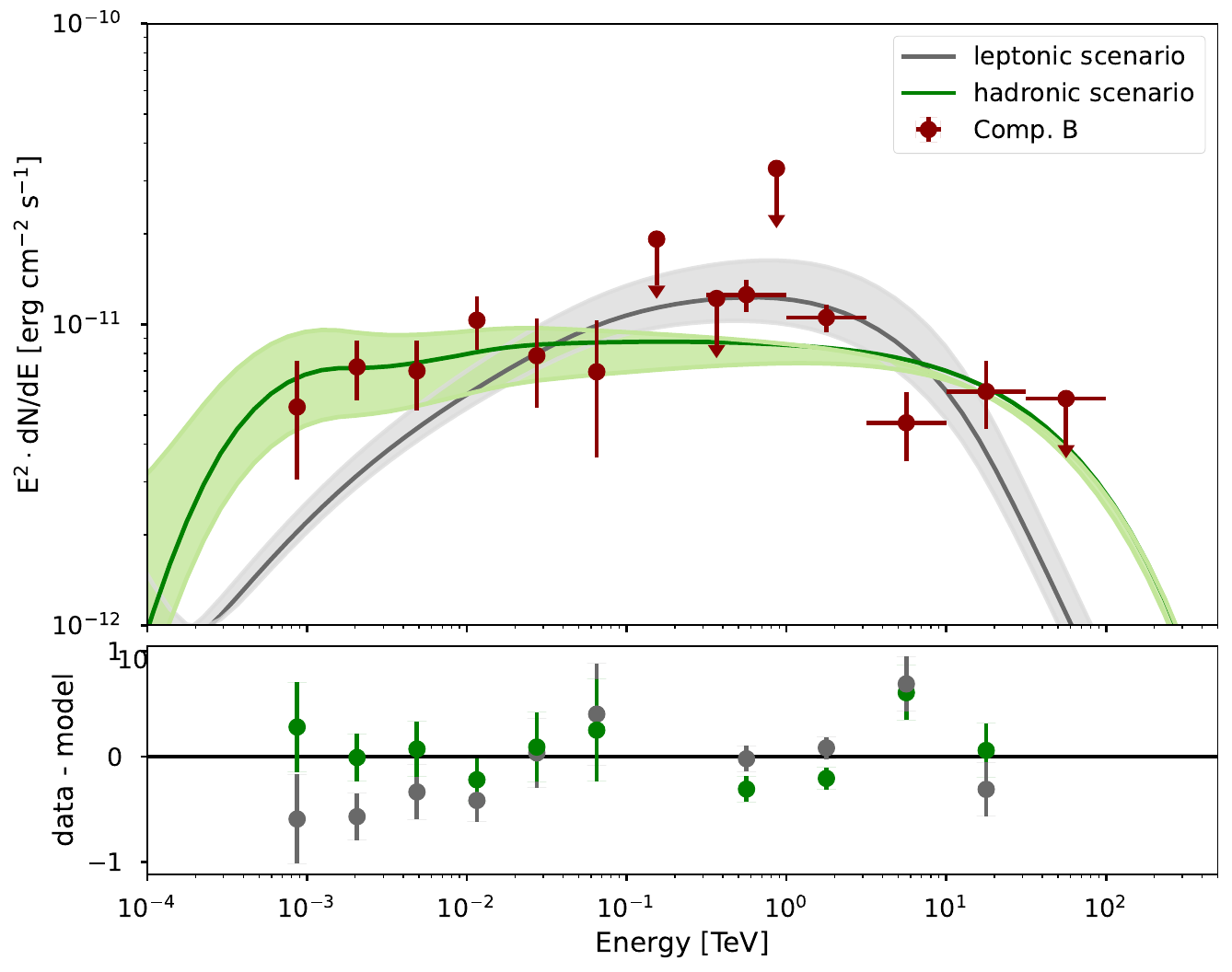}
\caption{The estimated $\gamma$-ray flux for the hadronic model compared to the $\gamma$-ray flux expected for the leptonic model computed in Section \ref{lept}. The deviation of the SED to the respective model curves are shown in the bottom panel.
}
\label{fig-modeling:hadronic}
\end{figure}

We evolve this particle population with time following the approach in Section \ref{lept}, using a MCMC-Chain. The fit parameters of the model, as well as the adjusted parameter range can be seen in table \ref{tab-modeling:fit-parameters_hadronic}. A comparison of the leptonic model derived in Section \ref{lept} and the hadronic model can be seen in Figure \ref{fig-modeling:hadronic}, while Figure \ref{fig-appendix:WDC_spec} shows the comparison of both models with the SED derived in this analysis, as well as the spectra derived from the emission observed in HAWC \citep{hawc_cat} and LHAASO \citep{lhaaso}.

Both models exhibit good agreement with the SED from component B as derived in section \ref{joint-analysis}, while only the leptonic model shows good agreement with the spectrum of 1LHAASO~J1814$-$1719u* measured by KM2A. The spectrum of 3HWC~J1813$-$174 does not show good agreement with either model.

Additionally to the discrepancy with the spectra derived by HAWC and LHAASO, the hadronic model has further caveats. The estimated distance between the molecular cloud at $4.6\,$kpc and the SNR at $6.2 - 12\,$kpc are not compatible, and the location of the molecular cloud can account for the emission observed in the GeV energy range, but not the extension of the TeV emssion. Assuming a particle density of $d = 1\,\text{cm}^{-3}$, corresponding to ISM level, however results in the need for a very powerful SNR with the energy of the protons required to be $\log_{10}(E_p/\text{erg}) = [51.2 - 51.7]\,$, to describe the emission well. Since there is no evidence supporting this scenario, a origin of the observed emission only from acceleration of protons on the SNR seems unlikely.

While such a high energy output seems unlikely in the case of only one object, collective effects of a population of young stars in the stellar cluster CL~J1813-178 (located at a distance of $4.8\,$kpc \citet{stellar_cluster}) could account for the required energy. For example, for a kinetic luminosity of $3\times10^{38}$\,erg/s and an age of 3-10\,Myr, as may be typical for a young, massive stellar cluster, the available total energy budget is $\log_{10}(E_p/\text{erg}) = [52.5 - 53]\,$ \citep{2021MNRAS.504.6096Morlino}.
\\
Acceleration of protons in the stellar cluster remains therefore a valid possible origin for the observed extended $\gamma$-ray emission, a detailed examination is, however, beyond the scope of this paper.

\begin{table}[H]
\caption{The validity range of the free parameters of the hadronic model.} 
\label{tab-modeling:fit-parameters_hadronic} 
\centering                          
\begin{tabular}{c | c  }       
\hline 
\hline 
\noalign{\smallskip}
    $\alpha_\text{p}$ & $[2.01 - 2.08]$ \\
    $\log_{10}(E_p)$ & $[48.49 - 49.65]\,\log{(\text{erg})}$ \\
    $\Delta \alpha$ & $-[0.15 - 0.27]\,$ \\
    $k_{ep}$ & $-[1.72 - 6.05]\,$ \\
    $B_\text{now}$ & $[73 - 271]\,\mu G$ \\
\noalign{\smallskip}
\hline    
\hline 
\end{tabular}
\end{table}


\section{Discussion}
In previous analyses of the region around \psr the analysis of H.E.S.S. data revealed compact emission with an extension of $0.06^\circ$ around the pulsar \citep{HGPS}, while the analysis of \fermilat data, as well as data aquired by HAWC and LHAASO showed largely extended emission. With these results, no connection between the \fermilat source \fermi and the H.E.S.S. source \source could be established. We reanalysed the region using an improved reconstruction for the TeV data and increased exposure in the GeV energy range. 

In the TeV energy range, we confirm the detection of \source, a bright, TeV $\gamma$-ray source, centred at the position of the \psr, with an extension of $0.06^\circ$, observed as HESS~J1813$-$178A in this analysis. The results derived for this emission component are consistent with the results derived in \citet{HGPS2}. We also detect a fainter $\gamma$-ray structure, with an extension of $0.7^\circ$ enclosing the pulsar and HESS~J1813$-$178A, which we refer to as HESS~J1813$-$178B. 

In the GeV energy range observed with \fermilat, we find $\gamma$-ray emission with an extension of $0.4^\circ$, positionally coincident with the extended emission observed in the H.E.S.S. dataset. Additionally, we observe compact emission positionally coincident with the pulsar. This emission has already been reported by \citet{fermi_araya}, but was not significant. While we also cannot claim a significant detection, adding component C improves the region's description. 

By combining the datasets in a joint analysis we find that the extended emission detected in the H.E.S.S. and \fermilat data can be connected and described by a single source model, while the emission from HESS~J1813$-$178A can only be observed in the H.E.S.S. data, since the emission drops below the sensitivity of the detector in the \fermilat energy range. We can therefore establish a consistent description of the region over five decades of energy and conclude that the emission observed by \fermilat and \hess have the same origin.

\citet{fermi_araya} concluded that the emission was most likely of hadronic origin due to the spectral shape. 
While \citet{fermi_araya} only took $\gamma$-rays in the GeV energy range into account, in this work, a combined model of GeV and TeV photons was used, as well as X-ray data included. We computed a physical model that can describe the observed $\gamma$-ray emission by evolving an electron population, or alternatively an proton population, over time. We find that the leptonic emission scenario describes the combined data well, while a acceleration at the SNR shock front seems less justified since the present target material cannot explain the observed extension of the TeV emission. The energetics necessary to produce the observed $\gamma$-ray emission could also be explained by acceleration of protons inside the stellar cluster, but has not been further investigated in this study.

We find, that the data from HESS~J1813$-$178A, as well as HESS~J1813$-$178B can be described well by $\gamma$-ray emission from synchrotron and IC emission of electrons originating from the pulsar. This indicates that HESS~J1813$-$178A, is likely a PWN, while the extended emission is possibly caused by electrons and positrons escaping the confines of the PWN and diffusing into the ISM. The effect of constructing a model that considers only time dependence, but not spatial dependence, leads to an overprediction in the keV energy range. 
Since the measurement of the flux observed by INTEGRAL and XMM-Newton only includes a small area (with an extension of less than $80"$), while the measurement of the flux from HESS~J1813$-$178B was conducted in an area of $\sim 0.5\degr$,  we can assume that a low surface brightness, as well as the limitations of the detectors cause the difference between predicted flux and measured flux in the X-ray energy range. 

Diffuse leptonic emission around a pulsar is usually only observed in systems older than $\sim 10\,$kyr \citep{pwn_population}. The true age estimated in this study, as well as previous observations of \psr \citep{quasars}, indicate that the system is younger than $5\,$kyrs, but it is not implausible that environmental factors, such as the local ISM density distribution or the pulsar proper motion, might lead to particles diffusing outwards faster than in previously observed systems. 

The estimated diffusion coefficient for HESS~J1813$-$178B is within the expected range of diffusion in the ISM \citep{ISM_diff} and comparable to the diffusion coefficient estimated for the $\gamma$-ray emission observed around PSR~B1823$-$13, the pulsar powering the highly extended PWN HESS~J1825$-$137 \citep{J1825}. 
Assuming the same diffusion coefficient for HESS~J1813$-$178A, would lead to a extension far bigger than the observed emission, implying that the diffusion coefficient is non uniform, which is expected for a pulsar environment. Additionally, we find that the energy density estimated for the extended emission is comparable to the densities estimated for the halos observed around the evolved systems around PSR~J0633$+$1746 and PSR~B0656$+$14 \citep{pwn_population}. While we would expect the diffusion coefficient in a TeV halo to be below the diffusion observed in the ISM, the low energy density and lack of other accelerators in the area suggest that the system around \psr shows characteristics of an older, evolved system despite its young age. Although it would be unusual to note electron escape from young PWNe, this may occur if the system becomes disrupted at an early stage, such as due to an early return of the supernova reverse shock preferentially from one direction. 

This study also investigates the possibility of a hadronic origin of the emission, by evolving a proton population originating from SNR G012.8–00.0.

In order to estimate the statistical description of both models, the absolute chi-squared ($\chi^2$) goodness-of-fit was estimated using the SED of component B derived in the joint-fit. This study finds $\chi^2_\text{had} = 18.56$ for the hadronic model and $\chi^2_\text{lep} = 30.62$ for the leptonic model. The hadronic model is therefore preferred at a $3.47\sigma$ level.

Whilst the hadronic model yields a marginally better statistical description for the emission, the target material identified in the region is not sufficient to account for the extent of the detected $\gamma$-ray emission and target material with a density of the level of the ISM would require an unreasonably high proton energy to explain the detected $\gamma$-ray flux.

For the data below $10\,$GeV, observed as component C, we cannot find evidence that convincingly points towards either a leptonic or hadronic origin. A hadronic scenario cannot be supported since no association to a nearby interstellar cloud or target material can be made. A leptonic scenario would imply the existence of a second electron population, or an unreasonably high particle density, which is only experienced by this second electron population. Another possible leptonic scenario might be a pulsar origin, which may alternatively account for the emission component, but this study cannot find convincing evidence for such an explanation. 
For a firm identification of the soft $\gamma$-ray emission as originating from the pulsar, a dedicated study of the emission at lower energies needs to be performed.


While we are not able to fully resolve the origin of the $\gamma$-ray emission in the region around \psr, we present a detailed morphological and spectral analysis of the region, as well as possible emission scenarios. To further add to the understanding about this source, more observations at high-energy $\gamma$-rays are necessary, especially the addition of further data taken by the Water Cherenkov detector arrays HAWC or LHAASO into the joint-fit and the addition of further data taken by \fermilat will be of great value in constraining the possible origin of the different components.

\begin{acknowledgements}
The support of the Namibian authorities and of the University of Namibia in facilitating the construction and operation of H.E.S.S. is gratefully acknowledged, as is the support by the German Ministry for Education and Research (BMBF), the Max Planck Society, the German Research Foundation (DFG), the Helmholtz Association, the Alexander von Humboldt Foundation, the French Ministry of Higher Education, Research and Innovation, the Centre National de la Recherche Scientifique (CNRS/IN2P3 and CNRS/INSU), the Commissariat à l’Énergie atomique et aux Énergies alternatives (CEA), the U.K. Science and Technology Facilities Council (STFC), the Irish Research Council (IRC) and the Science Foundation Ireland (SFI), the Knut and Alice Wallenberg Foundation, the Polish Ministry of Education and Science, agreement no. 2021/WK/06, the South African Department of Science and Technology and National Research Foundation, the University of Namibia, the National Commission on Research, Science \& Technology of Namibia (NCRST), the Austrian Federal Ministry of Education, Science and Research and the Austrian Science Fund (FWF), the Australian Research Council (ARC), the Japan Society for the Promotion of Science, the University of Amsterdam and the Science Committee of Armenia grant 21AG-1C085. We appreciate the excellent work of the technical support staff in Berlin, Zeuthen, Heidelberg, Palaiseau, Paris, Saclay, Tübingen and in Namibia in the construction and operation of the equipment. This work benefited from services provided by the H.E.S.S. Virtual Organisation, supported by the national resource providers of the EGI Federation.
\end{acknowledgements}

\bibliographystyle{aa} 
\bibliography{mybibliography.bib}

\begin{appendix} 

\section{Estimation of the hadronic background}\label{bkg}
The hadronic background was estimated using a template model constructed from archival H.E.S.S. data (see \citet{fov-bkgmodel}). To account for the variation in cosmic-ray flux for the respective observation conditions, a 3D likelihood fit of this template background model for every observation was performed, using the background amplitude $\Phi$ and spectral index $\delta$ as fit parameters. Using the spectral index small inacurracies in the spectral shape of the background model can be corrected by modifying the predicted background rate $R^*_\text{BG} = R_\text{BG} \cdot (E/E_0)^{- \delta_\text{BG}}$ at Energy $E$, with a reference energy $E_0 = 1\,$TeV.

\begin{figure}
\includegraphics[width=9cm]{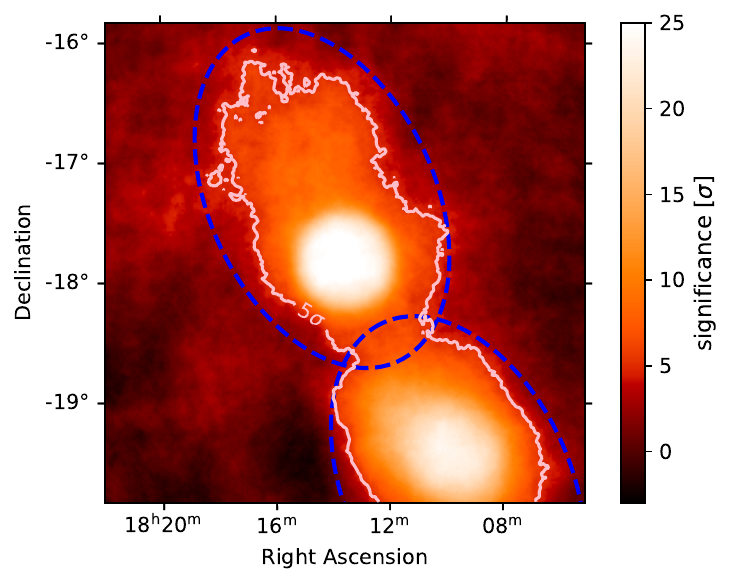}
\caption{Significance map of the region around \source, as seen after the fit of the background model, the exclusion regions used for the fitting are indicated by the dashed lines. A correlation radius of $0.4^\circ$ is used for the computation of this map.}
\label{fig-appendix:exclusion-regions}
\end{figure}

\begin{figure}
\includegraphics[width=9cm]{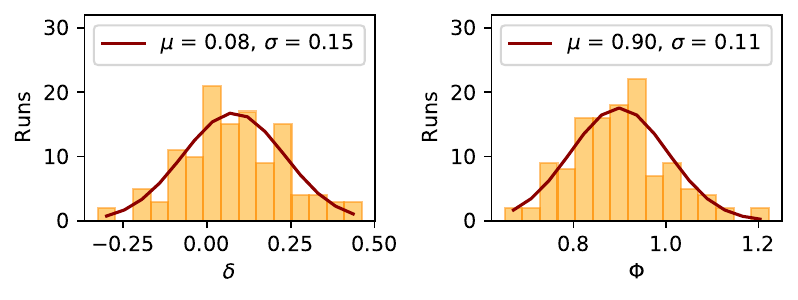}
\caption{The distribution of the fit parameters of the template background model and a Gaussian fit to the respective distribution. The mean and standard deviation of the background spectral index $\delta$ and amplitude $\Phi$ are indicated.}
\label{fig-appendix:bkg}
\end{figure}

\begin{figure}
\includegraphics[width=9cm]{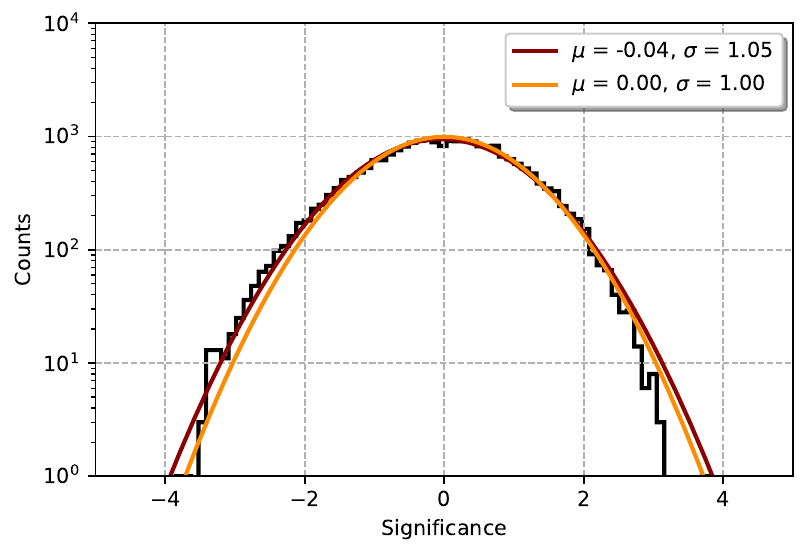}
\caption{Significance distribution of the data outside of the defined exclusion regions. The orange curve shows the expected behaviour for a normal distribution, indicating that only statistical fluctuations are present, while the red curve shows a fit of a Gaussian distribution to the data.}
\label{fig-appendix:sig_distribution}
\end{figure}

To get an accurate estimation of the background counts the regions in which significant emission is detected needs to be excluded from the background fit. These excluded regions are shown in Fig. \ref{fig-appendix:exclusion-regions}.
Figure \ref{fig-appendix:bkg} shows the distribution of the background amplitude and spectral index, as well as a Gaussian fit indicated in red, while Fig. \ref{fig-appendix:sig_distribution} shows the significance distribution outside of the exclusion regions.
This distribution follows a Gaussian distribution with a mean close two zero and a width that is close to one, indicating that only statistical fluctuations are present outside of the exclusion regions. 

\section{Calculation of systematic uncertainties}
\label{appendix:sys}
\begin{figure}
\includegraphics[width=9cm]{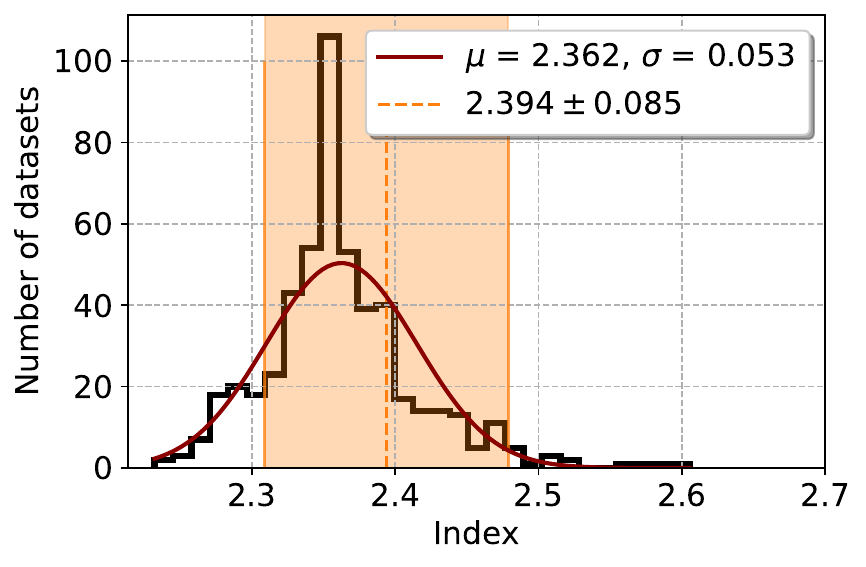}
\caption{Distribution of the best-fit values of the spectral index for HESS~J1813$-$178B. A Gaussian fit is performed on the data, which is indicated in red. The orange dashed line corresponds to the fit value derived in the analysis of the original dataset, the orange band indicates the statistical error.}
\label{fig-appendix:sys_errors}
\end{figure}

In order to derive the systematic uncertainties introduced by a mismodeling of the energy axis of the IRFs, a randomly generated value is drawn from a Gaussian distribution with a mean of 1 and a standard deviation of 10\%. We apply this factor $\varrho_E$ to the energy axes of the IRFs.
\\
To estimate the uncertainties introduced through mis-modeling of the hadronic background, a linear gradient with a direction angle $\alpha_{bkg}$ and a gradient amplitude $\Phi_{bkg}$ is applied to the background model. In addition to the gradient parameters, the amplitude $\Phi$ and index $\delta$ of the model is varied by randomly drawing a value from Gaussian distributions. The mean and deviation of these distributions are indicated in table \ref{tab-appendix:sys_params}.
\begin{table}
\caption{Parameter distribution used for the computation of the systematic errors.}
\label{tab-appendix:sys_params} 
\centering                          
\begin{tabular}{c | c c }       
\hline\hline  
\noalign{\smallskip}
parameter & distribution & variation  \\    
\noalign{\smallskip}
\hline 
\noalign{\smallskip}
   $\varrho_E$ & Gaussian & $\mu = 1, \sigma=0.1$  \\
   $\Phi$ & Gaussian & $\mu = 1, \sigma=0.05$   \\
   $\delta$ & Gaussian & $\mu = 0, \sigma=0.05$  \\
   $\alpha_{bkg}$ & uniform & $0^\circ - 360^\circ$  \\
   $\Phi_{bkg}$ & Gaussian & $\mu = 1, \sigma=0.01$  \\
\noalign{\smallskip}
\hline     
\hline 
\end{tabular}
\end{table}
The standard deviation of these distributions was chosen such that they represent the deviation between the mean value estimated in the pre-fit of the background model (described in Appendix \ref{bkg}) and the values estimated in the likelihood minimisation on the whole dataset. 
\\
Then, the shift, corresponding to the randomly drawn factor, was applied to the IRFs of the dataset and the background gradient introduced. The best-fit model, described in section \ref{result}, was applied, and the likelihood fit executed. This process was repeated 500 times.
\\
An example of the resulting distribution of fit parameters can be seen in Fig. \ref{fig-appendix:sys_errors}, where the statistical error is indicated by the orange band and the orange dashed line represents the best-fit value derived in the analysis of the original dataset. The standard deviation of this Gaussian distribution represents the systematic error derived by this method. The small systematic uncertainties estimated for the spectral index and morphological parameters indicate that this method does not take into account all possible sources of systematic uncertainties. A more general estimation of uncertainties on the spectral index can be found in \citet{sys_index}. 

\section{Coverage of the region using H.E.S.S.}
\begin{figure}
\includegraphics[width=9cm]{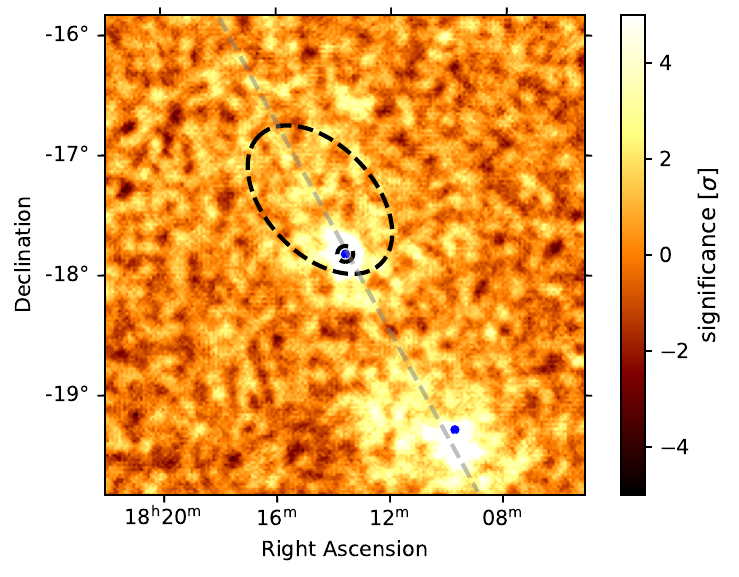}
\caption{Signficiance map of the ROI, computed with a correlation radius of $0.06^\circ$.}
\label{fig-appendix:sig_small_corr}
\end{figure}

\begin{figure}
\includegraphics[width=9cm]{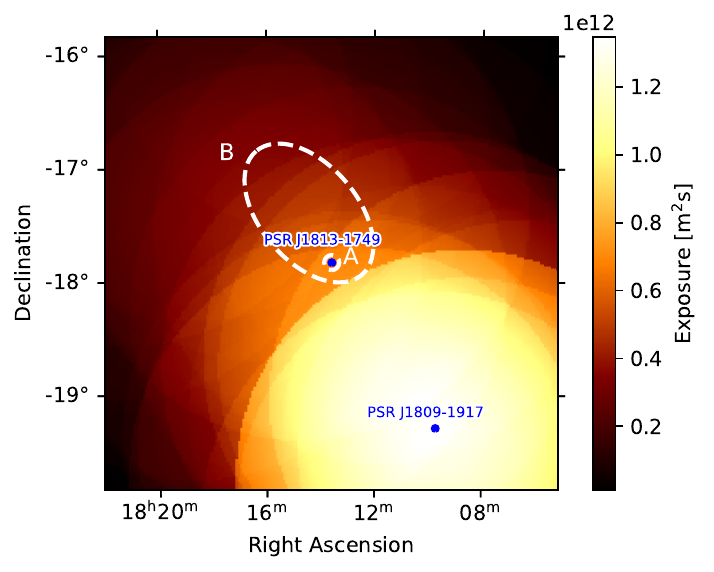}
\caption{Exposure map of the dataset used for the analysis of the H.E.S.S. data. Most of the exposure comes from observations centred on the neighbouring source $\text{HESS}\,\text{J}1809$-$193$.}
\label{fig-appendix:exposure}
\end{figure}

To visualize the coverage of the region using \hess, an exposure map (figure \ref{fig-appendix:exposure})
was computed. Additionally, a significance map computed with a small correlation radius of $0.06^\circ$, which is comparable to the PSF of the analysis, is depicted in figure \ref{fig-appendix:sig_small_corr}

\clearpage

\section{\emph{Fermi}-LAT data: Fermipy-Gammapy cross check}\label{export}
\begin{figure}
\includegraphics[width=9cm]{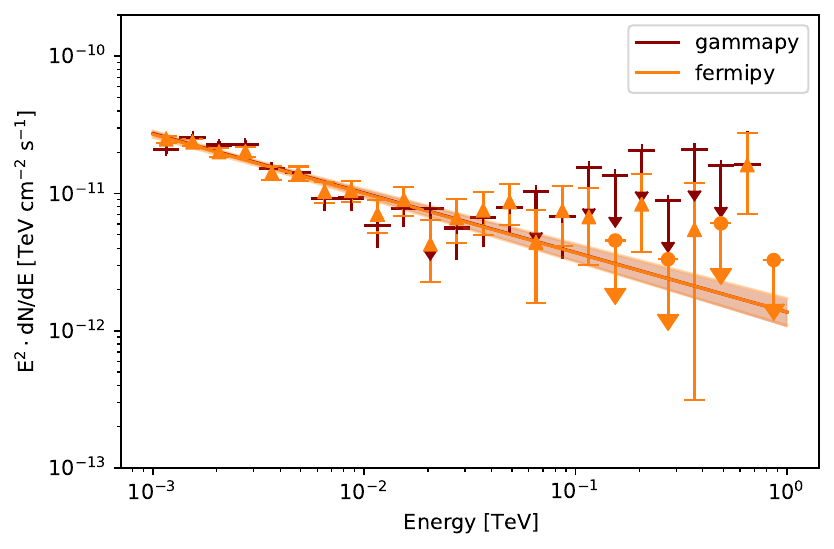}
\caption{ Best-fit spectrum of a power-law model applied to the \fermilat data in \texttt{fermipy} and \texttt{gammapy}.}

\label{fig-appendix:fermipy-gammapy}
\end{figure}

Whilst \texttt{gammapy} cannot be used to directly analyse \fermilat data, the \texttt{fermipy} analysis package is not able to perform a three-dimensional fit to the data. To combine both methods, the count's cube, the IRFs and the background model cubes were computed using the \texttt{fermipy} package. Then, the cubes and background source models were extracted and convert into a format that is supported by \texttt{gammapy}. The robustness of exporting the data and models to \texttt{gammapy} needs to be validated before the analysis results are used for further investigation. For this purpose, a simple analysis of the source region in \texttt{fermipy} and \texttt{gammapy} is carried out.
The morphology of the source given in the $4\text{FGL}$-catalog, a disc model with a source position of $\text{R.A.}=273.405\degr$, $\text{Dec.}=-17.653\degr$ and an extension of $0.6\degr$, was used for this validation. Then, the ROI was optimised with \texttt{fermipy} by only taking into account sources with a test statistics (TS) greater than four and predicted counts of more than 50. These cuts yielded a total of 24 source models present in the analysis region. The spectral parameters of these sources in a $5\degr$ radius around the catalogue position of \fermi were refitted. 

For the analysis in \texttt{gammapy}, the morphology of the source was fixed to the position derived in \texttt{fermipy} and the best-fit spectral parameters were evaluated. The spectral parameters and SED derived in \texttt{gammapy} and \texttt{fermipy} agree within errors. In addition to the spectral fit parameters, a comparison of the counts spectra of the background models and other source models in the ROI, can be seen in Appendix, Fig. \ref{fig-appendix:fermipy-gammapy}. We find that the results from both likelihood minimisations are in agreement with each other.

\clearpage

\section{Multiwavelength-context}
\begin{figure}[H]
\includegraphics[width=9cm]{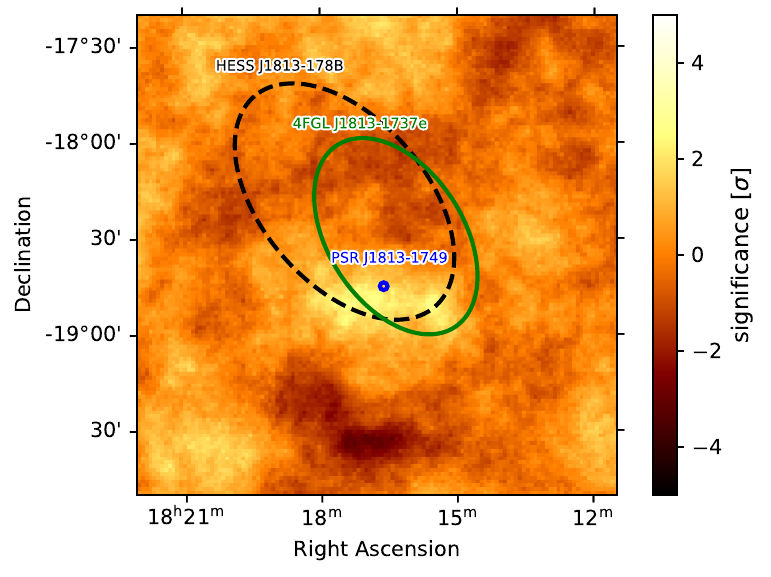}
\caption{Excess map of the H.E.S.S. data after the addition of the Gaussian and elongated Gaussian model. The position of \psr is indicated by the blue dot. The best-fit model describing the emission observed by H.E.S.S. is indicated by the black dashed lines. The \fermilat best-fit model is depicted by the green ellipse. The models show good spatial agreement}
\label{fig-appendix:combine_model}
\end{figure}

\begin{figure}
\includegraphics[width=9cm]{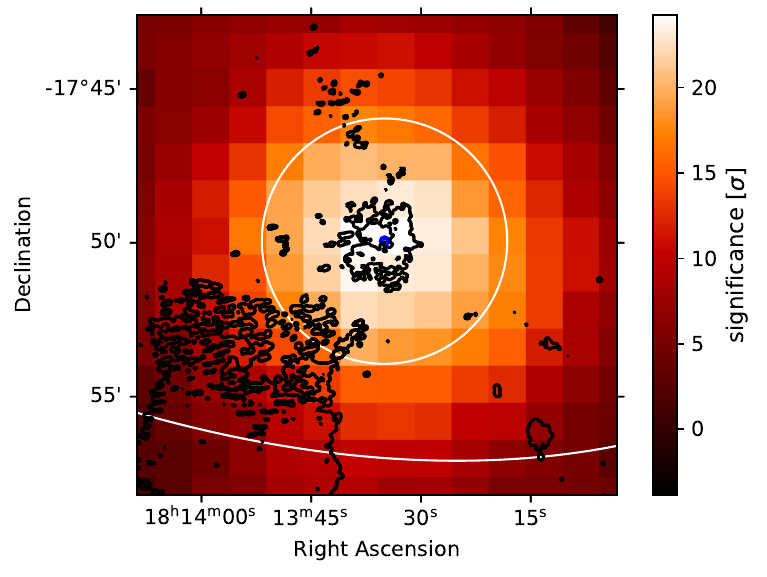}
\caption{Zoom-in on a significance map with a correlation radius of $0.06^\circ$ of the region around \psr as seen with \hess. The best-fit morphology of HESS~J1813$-$178A and HESS~J1813$-$178B are indicated by the white lines, the black countours depict SNR G012.8–00.0 and CL~J1813$-$178 as observed by VLA. The positon of \psr is indicated in blue.}
\label{fig-appendix:multiwavelenght}
\end{figure}

\begin{figure}
\includegraphics[width=9cm]{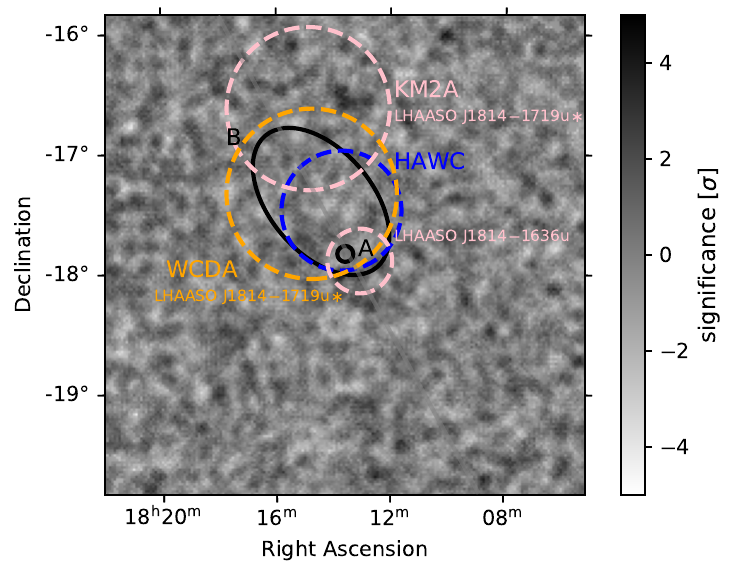}
\caption{The best-fit morphology derived from the analysis of the H.E.S.S. data is indicated in black. The best-fit morphology of the source 3HWC~J1813$-$174 is indicated in blue. The description of the region obtained with LHAASO is indicate by the pink and orange dashed circles.}
\label{fig-appendix:WDC_morphology}
\end{figure}

\begin{figure}
\includegraphics[width=9cm]{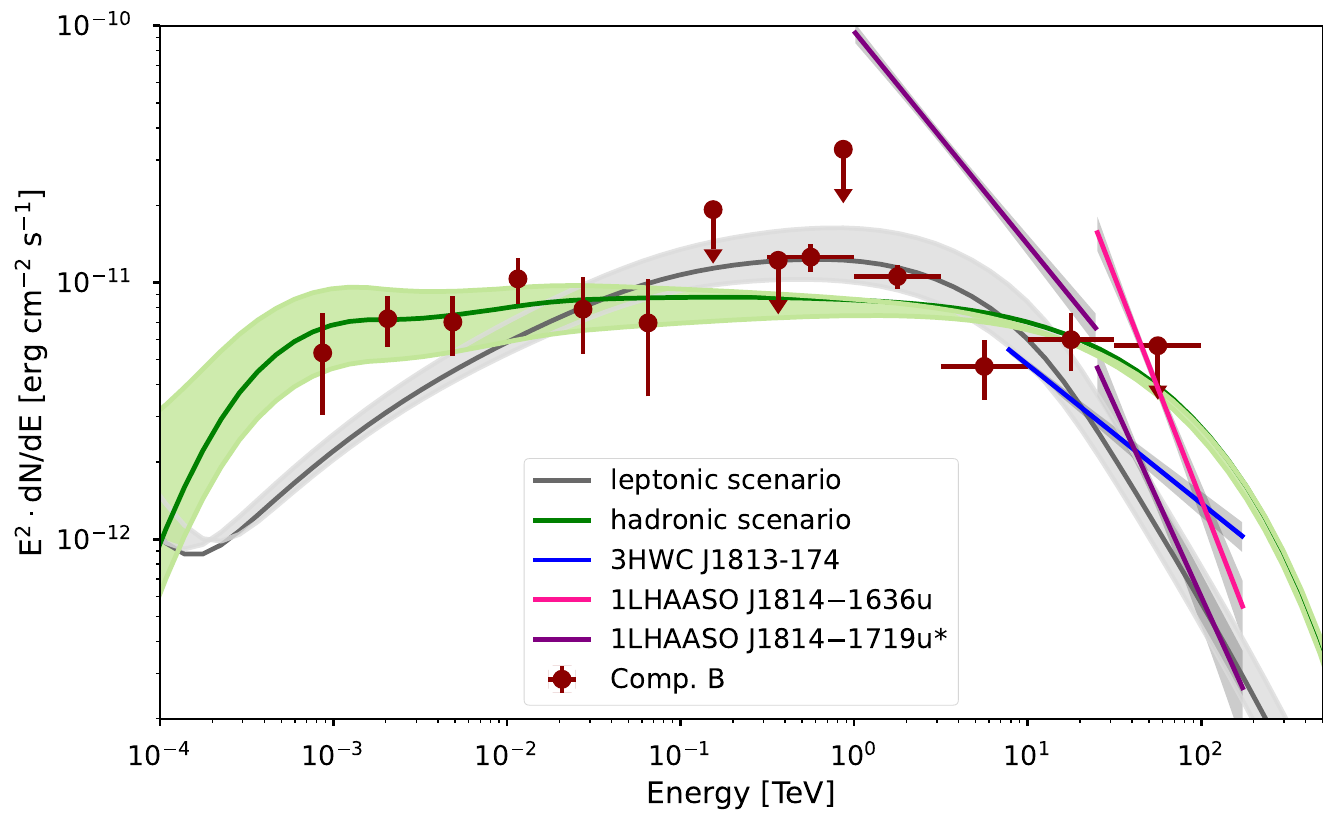}
\caption{The estimated $\gamma$-ray flux for the hadronic and leptonic model together with the spectra derived in the analysis of the Fermi-LAT, H.E.S.S., HAWC and LHAASO data.}
\label{fig-appendix:WDC_spec}
\end{figure}

To better understand the nature of the emission in the region around \psr, a comparison of the observed emission in different energy ranges was made. Comparing the best-fit model derived in the analysis of the \fermilat data and the H.E.S.S. data yields good spatial agreement (Fig. \ref{fig-appendix:combine_model}). The elongation is in both cases observed along the galactic plane and the pulsar is contained. 
Taking into account data taken by the VLA (red contours in Fig. \ref{fig-appendix:multiwavelenght}), a positional coincidence between HESS~J1813$-$178A, the pulsar and supernova remnant can be observed. Additionally shown in the figure is a part of the stellar cluster Cl~J1813$-$178 located in close vincinity to the pulsar.  
The region has also been observed by HAWC \citep{hawc_cat} and LHAASO \citep{lhaaso}, both detecting extended emission around the position of HESS~J1813$-$178A. In order to obtain a better understanding of the region, the position and extension of these sources, as well as the best-fit model obtained in this work, are visualized in a significance map computed from the H.E.S.S. data (see figure \ref{fig-appendix:WDC_morphology}). 
The morphology of the different components show a good spatial agreement. Additionally to the morphology, the spectra, derived in \citet{hawc_cat} and \citet{lhaaso} are visualized in figure \ref{fig-appendix:WDC_spec}, together with the leptonic and hadronic model derived in this work. The spectra of 1LHAASO~J1814$-$1719u*, observed by KM2A and 3HWC~J1813$-$174 show a good agreement with the results derived in this study, while the spectrum of 1LHAASO~J1814$-$1719u*, observed by WCDA as well as the spectrum of 1LHAASO~J1814$-$1636u do not agree. A possible reason for this could be the differences in the separation between source emission and diffuse emission along the galactic plane.

\end{appendix}

\end{document}

%% file: authors.tex
\author{H.E.S.S. Collaboration
\and F.~Aharonian \inst{\ref{DIAS},\ref{MPIK},\ref{Yerevan}}
\and F.~Ait~Benkhali \inst{\ref{LSW}}
\and J.~Aschersleben \inst{\ref{Groningen}}
\and H.~Ashkar \inst{\ref{LLR}}
\and M.~Backes \inst{\ref{UNAM},\ref{NWU}}
\and A.~Baktash \inst{\ref{UHH}}
\and V.~Barbosa~Martins \inst{\ref{DESY}}
\and J.~Barnard \inst{\ref{UFS}}
\and R.~Batzofin \inst{\ref{UP}}
\and Y.~Becherini \inst{\ref{APC},\ref{Linnaeus}}
\and D.~Berge \inst{\ref{DESY},\ref{HUB}}
\and K.~Bernl\"ohr \inst{\ref{MPIK}}
\and B.~Bi \inst{\ref{IAAT}}
\and M.~B\"ottcher \inst{\ref{NWU}}
\and C.~Boisson \inst{\ref{LUTH}}
\and J.~Bolmont \inst{\ref{LPNHE}}
\and M.~de~Bony~de~Lavergne \inst{\ref{CEA}}
\and J.~Borowska \inst{\ref{HUB}}
\and M.~Bouyahiaoui \inst{\ref{MPIK}}
\and M.~Breuhaus \inst{\ref{MPIK}}
\and R.~Brose \inst{\ref{DIAS}}
\and F.~Brun \inst{\ref{CEA}}
\and B.~Bruno \inst{\ref{ECAP}}
\and T.~Bulik \inst{\ref{UWarsaw}}
\and C.~Burger-Scheidlin \inst{\ref{DIAS}}
\and S.~Caroff \inst{\ref{LAPP}}
\and S.~Casanova \inst{\ref{IFJPAN}}
\and R.~Cecil \inst{\ref{UHH}}
\and J.~Celic \inst{\ref{ECAP}}
\and M.~Cerruti \inst{\ref{APC}}
\and P.~Chambery \inst{\ref{CENBG}}\footnotemark[1]
\and T.~Chand \inst{\ref{NWU}}
\and A.~Chen \inst{\ref{Wits}}
\and J.~Chibueze \inst{\ref{NWU}}
\and O.~Chibueze \inst{\ref{NWU}}
\and G.~Cotter \inst{\ref{Oxford}}
\and J.~Damascene~Mbarubucyeye \inst{\ref{DESY}}
\and A.~Djannati-Ata\"i \inst{\ref{APC}}
\and A.~Dmytriiev \inst{\ref{NWU}}
\and V.~Doroshenko \inst{\ref{IAAT}}
\and S.~Einecke \inst{\ref{Adelaide}}
\and J.-P.~Ernenwein \inst{\ref{CPPM}}
\and K.~Feijen \inst{\ref{APC}}
\and M.~Filipovic \inst{\ref{WSU}}
\and G.~Fontaine \inst{\ref{LLR}}
\and M.~F\"u{\ss}ling \inst{\ref{DESY}}
\and S.~Funk \inst{\ref{ECAP}}
\and S.~Gabici \inst{\ref{APC}}
\and Y.A.~Gallant \inst{\ref{LUPM}}
\and S.~Ghafourizadeh \inst{\ref{LSW}}
\and G.~Giavitto \inst{\ref{DESY}}
\and D.~Glawion \inst{\ref{ECAP}}
\and J.F.~Glicenstein \inst{\ref{CEA}}
\and P.~Goswami \inst{\ref{APC}}
\and G.~Grolleron \inst{\ref{LPNHE}}
\and M.-H.~Grondin \inst{\ref{CENBG}}
\and J.A.~Hinton \inst{\ref{MPIK}}
\and W.~Hofmann \inst{\ref{MPIK}}
\and T.~L.~Holch \inst{\ref{DESY}}
\and M.~Holler \inst{\ref{Innsbruck}}
\and M.~Jamrozy \inst{\ref{UJK}}
\and F.~Jankowsky \inst{\ref{LSW}}
\and V.~Joshi \inst{\ref{ECAP}}\footnotemark[1]
\and I.~Jung-Richardt \inst{\ref{ECAP}}
\and K.~Katarzy{\'n}ski \inst{\ref{NCUT}}
\and R.~Khatoon \inst{\ref{NWU}}
\and B.~Kh\'elifi \inst{\ref{APC}}
\and S.~Klepser \inst{\ref{DESY}}
\and W.~Klu\'{z}niak \inst{\ref{NCAC}}
\and Nu.~Komin \inst{\ref{Wits}}
\and K.~Kosack \inst{\ref{CEA}}
\and D.~Kostunin \inst{\ref{DESY}}
\and A.~Kundu \inst{\ref{NWU}}
\and R.G.~Lang \inst{\ref{ECAP}}
\and S.~Le~Stum \inst{\ref{CPPM}}
\and F.~Leitl \inst{\ref{ECAP}}
\and A.~Lemi\`ere \inst{\ref{APC}}
\and M.~Lemoine-Goumard \inst{\ref{CENBG}}
\and J.-P.~Lenain \inst{\ref{LPNHE}}
\and F.~Leuschner \inst{\ref{IAAT}}
\and J.~Mackey \inst{\ref{DIAS}}
\and D.~Malyshev \inst{\ref{IAAT}}
\and D.~Malyshev \inst{\ref{ECAP}}
\and V.~Marandon \inst{\ref{CEA}}
\and P.~Marinos \inst{\ref{Adelaide}}
\and G.~Mart\'i-Devesa \inst{\ref{Innsbruck}}
\and R.~Marx \inst{\ref{LSW}}
\and A.~Mehta \inst{\ref{DESY}}
\and A.~Mitchell \inst{\ref{ECAP}}\footnotemark[1]
\and R.~Moderski \inst{\ref{NCAC}}
\and L.~Mohrmann \inst{\ref{MPIK}}
\and A.~Montanari \inst{\ref{LSW}}
\and E.~Moulin \inst{\ref{CEA}}
\and T.~Murach \inst{\ref{DESY}}
\and M.~de~Naurois \inst{\ref{LLR}}
\and J.~Niemiec \inst{\ref{IFJPAN}}
\and A.~Priyana~Noel \inst{\ref{UJK}}
\and P.~O'Brien \inst{\ref{Leicester}}
\and S.~Ohm \inst{\ref{DESY}}
\and L.~Olivera-Nieto \inst{\ref{MPIK}}
\and E.~de~Ona~Wilhelmi \inst{\ref{DESY}}
\and M.~Ostrowski \inst{\ref{UJK}}
\and S.~Panny \inst{\ref{Innsbruck}}
\and M.~Panter \inst{\ref{MPIK}}
\and R.D.~Parsons \inst{\ref{HUB}}
\and D.A.~Prokhorov \inst{\ref{Amsterdam}}
\and G.~P\"uhlhofer \inst{\ref{IAAT}}
\and M.~Punch \inst{\ref{APC}}
\and A.~Quirrenbach \inst{\ref{LSW}}
\and M.~Regeard \inst{\ref{APC}}
\and P.~Reichherzer \inst{\ref{CEA}}
\and A.~Reimer \inst{\ref{Innsbruck}}
\and O.~Reimer \inst{\ref{Innsbruck}}
\and H.~Ren \inst{\ref{MPIK}}
\and M.~Renaud \inst{\ref{LUPM}}
\and B.~Reville \inst{\ref{MPIK}}
\and F.~Rieger \inst{\ref{MPIK}}
\and G.~Roellinghoff \inst{\ref{ECAP}}
\and B.~Rudak \inst{\ref{NCAC}}
\and V.~Sahakian \inst{\ref{Yerevan2}}
\and H.~Salzmann \inst{\ref{IAAT}}
\and M.~Sasaki \inst{\ref{ECAP}}
\and F.~Sch\"ussler \inst{\ref{CEA}}
\and H.M.~Schutte \inst{\ref{NWU}}
\and J.N.S.~Shapopi \inst{\ref{UNAM}}
\and A.~Specovius \inst{\ref{ECAP}}
\and S.~Spencer \inst{\ref{ECAP}}
\and R.~Steenkamp \inst{\ref{UNAM}}
\and S.~Steinmassl \inst{\ref{MPIK}}
\and C.~Steppa \inst{\ref{UP}}
\and I.~Sushch \inst{\ref{NWU}}
\and H.~Suzuki \inst{\ref{Konan}}
\and T.~Takahashi \inst{\ref{KAVLI}}
\and T.~Tanaka \inst{\ref{Konan}}
\and R.~Terrier \inst{\ref{APC}}
\and M.~Tluczykont \inst{\ref{UHH}}
\and N.~Tsuji \inst{\ref{RIKKEN}}
\and T.~Unbehaun \inst{\ref{ECAP}}
\and C.~van~Eldik \inst{\ref{ECAP}}
\and M.~Vecchi \inst{\ref{Groningen}}
\and J.~Veh \inst{\ref{ECAP}}
\and C.~Venter \inst{\ref{NWU}}
\and J.~Vink \inst{\ref{Amsterdam}}
\and T.~Wach \inst{\ref{ECAP}}\thanks{Corresponding authors;\\
e-mail: contact.hess@hess-experiment.eu}
\and S.J.~Wagner \inst{\ref{LSW}}
\and A.~Wierzcholska \inst{\ref{IFJPAN}}
\and M.~Zacharias \inst{\ref{LSW},\ref{NWU}}
\and D.~Zargaryan \inst{\ref{DIAS}}
\and A.A.~Zdziarski \inst{\ref{NCAC}}
\and S.~Zouari \inst{\ref{APC}}
\and N.~\.Zywucka \inst{\ref{NWU}}
}

\institute{
Dublin Institute for Advanced Studies, 31 Fitzwilliam Place, Dublin 2, Ireland \label{DIAS} \and
Max-Planck-Institut f\"ur Kernphysik, P.O. Box 103980, D 69029 Heidelberg, Germany \label{MPIK} \and
Yerevan Physics Institute, 2 Alikhanian Brothers St., Yerevan 0036, Armenia \label{Yerevan2} \and
Yerevan State University,  1 Alek Manukyan St, Yerevan 0025, Armenia \label{Yerevan} \and
Landessternwarte, Universit\"at Heidelberg, K\"onigstuhl, D 69117 Heidelberg, Germany \label{LSW} \and
Kapteyn Astronomical Institute, University of Groningen, Landleven 12, 9747 AD Groningen, The Netherlands \label{Groningen} \and
Laboratoire Leprince-Ringuet, École Polytechnique, CNRS, Institut Polytechnique de Paris, F-91128 Palaiseau, France \label{LLR} \and
University of Namibia, Department of Physics, Private Bag 13301, Windhoek 10005, Namibia \label{UNAM} \and
Centre for Space Research, North-West University, Potchefstroom 2520, South Africa \label{NWU} \and
Universit\"at Hamburg, Institut f\"ur Experimentalphysik, Luruper Chaussee 149, D 22761 Hamburg, Germany \label{UHH} \and
Deutsches Elektronen-Synchrotron DESY, Platanenallee 6, 15738 Zeuthen, Germany \label{DESY} \and
Department of Physics, University of the Free State,  PO Box 339, Bloemfontein 9300, South Africa \label{UFS} \and
Institut f\"ur Physik und Astronomie, Universit\"at Potsdam,  Karl-Liebknecht-Strasse 24/25, D 14476 Potsdam, Germany \label{UP} \and
Université Paris Cité, CNRS, Astroparticule et Cosmologie, F-75013 Paris, France \label{APC} \and
Department of Physics and Electrical Engineering, Linnaeus University,  351 95 V\"axj\"o, Sweden \label{Linnaeus} \and
Institut f\"ur Physik, Humboldt-Universit\"at zu Berlin, Newtonstr. 15, D 12489 Berlin, Germany \label{HUB} \and
Institut f\"ur Astronomie und Astrophysik, Universit\"at T\"ubingen, Sand 1, D 72076 T\"ubingen, Germany \label{IAAT} \and
Laboratoire Univers et Théories, Observatoire de Paris, Université PSL, CNRS, Université Paris Cité, 5 Pl. Jules Janssen, 92190 Meudon, France \label{LUTH} \and
Sorbonne Universit\'e, Universit\'e Paris Diderot, Sorbonne Paris Cit\'e, CNRS/IN2P3, Laboratoire de Physique Nucl\'eaire et de Hautes Energies, LPNHE, 4 Place Jussieu, F-75252 Paris, France \label{LPNHE} \and
IRFU, CEA, Universit\'e Paris-Saclay, F-91191 Gif-sur-Yvette, France \label{CEA} \and
Friedrich-Alexander-Universit\"at Erlangen-N\"urnberg, Erlangen Centre for Astroparticle Physics, Nikolaus-Fiebiger-Str. 2, 91058 Erlangen, Germany \label{ECAP} \and
Astronomical Observatory, The University of Warsaw, Al. Ujazdowskie 4, 00-478 Warsaw, Poland \label{UWarsaw} \and
Université Savoie Mont Blanc, CNRS, Laboratoire d'Annecy de Physique des Particules - IN2P3, 74000 Annecy, France \label{LAPP} \and
Instytut Fizyki J\c{a}drowej PAN, ul. Radzikowskiego 152, 31-342 Krak{\'o}w, Poland \label{IFJPAN} \and
Universit\'e Bordeaux, CNRS, LP2I Bordeaux, UMR 5797, F-33170 Gradignan, France \label{CENBG} \and
School of Physics, University of the Witwatersrand, 1 Jan Smuts Avenue, Braamfontein, Johannesburg, 2050 South Africa \label{Wits} \and
University of Oxford, Department of Physics, Denys Wilkinson Building, Keble Road, Oxford OX1 3RH, UK \label{Oxford} \and
School of Physical Sciences, University of Adelaide, Adelaide 5005, Australia \label{Adelaide} \and
Aix Marseille Universit\'e, CNRS/IN2P3, CPPM, Marseille, France \label{CPPM} \and
Laboratoire Univers et Particules de Montpellier, Universit\'e Montpellier, CNRS/IN2P3,  CC 72, Place Eug\`ene Bataillon, F-34095 Montpellier Cedex 5, France \label{LUPM} \and
Universit\"at Innsbruck, Institut f\"ur Astro- und Teilchenphysik, Technikerstraße 25, 6020 Innsbruck, Austria \label{Innsbruck} \and
Obserwatorium Astronomiczne, Uniwersytet Jagiello{\'n}ski, ul. Orla 171, 30-244 Krak{\'o}w, Poland \label{UJK} \and
Institute of Astronomy, Faculty of Physics, Astronomy and Informatics, Nicolaus Copernicus University,  Grudziadzka 5, 87-100 Torun, Poland \label{NCUT} 
\newpage
\and Nicolaus Copernicus Astronomical Center, Polish Academy of Sciences, ul. Bartycka 18, 00-716 Warsaw, Poland \label{NCAC} \and
Department of Physics and Astronomy, The University of Leicester, University Road, Leicester, LE1 7RH, United Kingdom \label{Leicester} \and
GRAPPA, Anton Pannekoek Institute for Astronomy, University of Amsterdam,  Science Park 904, 1098 XH Amsterdam, The Netherlands \label{Amsterdam} \and
Department of Physics, Konan University, 8-9-1 Okamoto, Higashinada, Kobe, Hyogo 658-8501, Japan \label{Konan} \and
Kavli Institute for the Physics and Mathematics of the Universe (WPI), The University of Tokyo Institutes for Advanced Study (UTIAS), The University of Tokyo, 5-1-5 Kashiwa-no-Ha, Kashiwa, Chiba, 277-8583, Japan \label{KAVLI} \and
RIKEN, 2-1 Hirosawa, Wako, Saitama 351-0198, Japan \label{RIKKEN}
\and School of Science, Western Sydney University, Locked Bag 1797, Penrith South DC, NSW 2751, Australia \label{WSU}
}

%% file: aanda.bbl
\begin{thebibliography}{49}
\expandafter\ifx\csname natexlab\endcsname\relax\def\natexlab#1{#1}\fi

\bibitem[{{Abdalla} {et~al.}(2021){Abdalla}, {Aharonian}, {Ait Benkhali},
  {Ang{\"u}ner}, {Arcaro}, {Armand}, {Armstrong}, {Ashkar}, {Backes},
  {Baghmanyan}, {Barbosa Martins}, {Barnacka}, {Barnard}, {Becherini}, {Berge},
  {Bernl{\"o}hr}, {Bi}, {B{\"o}ttcher}, {Boisson}, {Bolmont}, {de Bonyde
  Lavergne}, {Breuhaus}, {Brose}, {Brun}, {Brun}, {Bryan}, {B{\"u}chele},
  {Bulik}, {Bylund}, {Caroff}, {Carosi}, {Chand}, {Chandra}, {Chen}, {Cotter},
  {Cury{\l}o}, {Damascene Mbarubucyeye}, {Davids}, {Davies}, {Deil}, {Devin},
  {Dirson}, {Djannati-Ata{\"\i}}, {Dmytriiev}, {Donath}, {Doroshenko},
  {Dreyer}, {Duffy}, {Dyks}, {Egberts}, {Eichhorn}, {Einecke}, {Emery},
  {Ernenwein}, {Feijen}, {Fegan}, {Fiasson}, {Fichet de Clairfontaine},
  {Fontaine}, {Funk}, {F{\"u}{\ss}ling}, {Gabici}, {Gallant}, {Giavitto},
  {Giunti}, {Glawion}, {Glicenstein}, {Gottschall}, {Grondin}, {Hahn}, {Haupt},
  {Hermann}, {Hinton}, {Hofmann}, {Hoischen}, {Holch}, {Holler}, {H{\"o}rbe},
  {Horns}, {Huber}, {Jamrozy}, {Jankowsky}, {Jankowsky}, {Jung-Richardt},
  {Kasai}, {Kastendieck}, {Katarzy{\'n}ski}, {Katz}, {Khangulyan},
  {Kh{\'e}lifi}, {Klepser}, {Klu{\'z}niak}, {Komin}, {Konno}, {Kosack},
  {Kostunin}, {Kreter}, {Lamanna}, {Lemi{\`e}re}, {Lemoine-Goumard}, {Lenain},
  {Leuschner}, {Levy}, {Lohse}, {Lypova}, {Mackey}, {Majumdar}, {Malyshev},
  {Malyshev}, {Marandon}, {Marchegiani}, {Marcowith}, {Mares},
  {Mart{\'\i}-Devesa}, {Marx}, {Maurin}, {Meintjes}, {Meyer}, {Mitchell},
  {Moderski}, {Mohrmann}, {Montanari}, {Moore}, {Morris}, {Moulin}, {Muller},
  {Murach}, {Nakashima}, {Nayerhoda}, {de Naurois}, {Ndiyavala}, {Niemiec},
  {Oakes}, {O'Brien}, {Odaka}, {Ohm}, {Olivera-Nieto}, {de Ona Wilhelmi},
  {Ostrowski}, {Panter}, {Panny}, {Parsons}, {Peron}, {Peyaud}, {Piel}, {Pita},
  {Poireau}, {Priyana Noel}, {Prokhorov}, {Prokoph}, {P{\"u}hlhofer}, {Punch},
  {Quirrenbach}, {Raab}, {Rauth}, {Reichherzer}, {Reimer}, {Reimer}, {Remy},
  {Renaud}, {Rieger}, {Rinchiuso}, {Romoli}, {Rowell}, {Rudak}, {Sahakian},
  {Sailer}, {Salzmann}, {Sanchez}, {Santangelo}, {Sasaki}, {Sch{\"a}fer},
  {Sch{\"u}ssler}, {Schutte}, {Schwanke}, {Seglar-Arroyo}, {Senniappan},
  {Seyffert}, {Shafi}, {Shapopi}, {Shiningayamwe}, {Simoni}, {Sinha}, {Sol},
  {Specovius}, {Spencer}, {Spir-Jacob}, {Stawarz}, {Sun}, {Steenkamp},
  {Stegmann}, {Steinmassl}, {Steppa}, {Takahashi}, {Tavernier}, {Taylor},
  {Terrier}, {Thiersen}, {Tiziani}, {Tluczykont}, {Tomankova}, {Trichard},
  {Tsirou}, {Tuffs}, {Uchiyama}, {van der Walt}, {van Eldik}, {van Rensburg},
  {van Soelen}, {Vasileiadis}, {Veh}, {Venter}, {Vincent}, {Vink}, {V{\"o}lk},
  {Wadiasingh}, {Wagner}, {Watson}, {Werner}, {White}, {Wierzcholska}, {Wong},
  {Yusafzai}, {Zacharias}, {Zanin}, {Zargaryan}, {Zdziarski}, {Zech}, {Zhu},
  {Zmija}, {Zorn}, {Zouari}, {{\.Z}ywucka}, {Albert}, {Alfaro}, {Alvarez},
  {Arteaga-Vel{\'e}zquez}, {Arunbabu}, {Avila Rojas}, {Belmont-Moreno},
  {BenZvi}, {Brisbois}, {Caballero-Mora}, {Capistr{\'a}n}, {Carrami{\~n}ana},
  {Casanova}, {Cotti}, {Cotzomi}, {Couti{\~n}o de Le{\'o}n}, {De la Fuente},
  {de Le{\'o}n}, {Diaz Hernandez}, {D{\'\i}az-V{\'e}lez}, {Dingus},
  {DuVernois}, {Durocher}, {Ellsworth}, {Engel}, {Espinoza}, {Fan},
  {Fern{\'a}ndez Alonso}, {Fraija}, {Galv{\'a}n-G{\'a}mez}, {Garcia},
  {Garc{\'\i}a-Gonz{\'a}lez}, {Garfias}, {Giacinti}, {Gonz{\'a}lez}, {Goodman},
  {Harding}, {Hernandez}, {Hona}, {Huang}, {Hueyotl-Zahuantitla},
  {H{\"u}ntemeyer}, {Iriarte}, {Jardin-Blicq}, {Joshi}, {Kieda}, {Lee},
  {Le{\'o}n Vargas}, {Linnemann}, {Longinotti}, {Luis-Raya}, {L{\'o}pez-Coto},
  {Malone}, {Martinez}, {Martinez-Castellanos}, {Mart{\'\i}nez-Castro},
  {Matthews}, {Miranda-Romagnoli}, {Morales-Soto}, {Moreno}, {Mostaf{\'a}},
  {Nayerhoda}, {Nellen}, {Newbold}, {Nisa}, {Noriega-Papaqui}, {Omodei},
  {Peisker}, {P{\'e}rez Araujo}, {P{\'e}rez-P{\'e}rez}, {Rho},
  {Rosa-Gonz{\'a}lez}, {Ruiz-Velasco}, {Salesa Greus}, {Sandoval}, {Schneider},
  {Schoorlemmer}, {Serna-Franco}, {Smith}, {Springer}, {Surajbali},
  {Tollefson}, {Torres}, {Torres-Escobedo}, {Turner}, {Ure{\~n}a-Mena},
  {Villase{\~n}or}, {Weisgarber}, {Willox}, {Zhou}, \& {HAWC
  Collaboration}}]{halo1}
{Abdalla}, H., {Aharonian}, F., {Ait Benkhali}, F., {et~al.} 2021, \apj, 917, 6

\bibitem[{{Abdollahi} {et~al.}(2020){Abdollahi}, {Acero}, {Ackermann},
  {Ajello}, {Atwood}, {Axelsson}, {Baldini}, {Ballet}, {Barbiellini},
  {Bastieri}, {Becerra Gonzalez}, {Bellazzini}, {Berretta}, {Bissaldi},
  {Blandford}, {Bloom}, {Bonino}, {Bottacini}, {Brandt}, {Bregeon}, {Bruel},
  {Buehler}, {Burnett}, {Buson}, {Cameron}, {Caputo}, {Caraveo}, {Casandjian},
  {Castro}, {Cavazzuti}, {Charles}, {Chaty}, {Chen}, {Cheung}, {Chiaro},
  {Ciprini}, {Cohen-Tanugi}, {Cominsky}, {Coronado-Bl{\'a}zquez}, {Costantin},
  {Cuoco}, {Cutini}, {D'Ammando}, {DeKlotz}, {de la Torre Luque}, {de Palma},
  {Desai}, {Digel}, {Di Lalla}, {Di Mauro}, {Di Venere}, {Dom{\'\i}nguez},
  {Dumora}, {Fana Dirirsa}, {Fegan}, {Ferrara}, {Franckowiak}, {Fukazawa},
  {Funk}, {Fusco}, {Gargano}, {Gasparrini}, {Giglietto}, {Giommi}, {Giordano},
  {Giroletti}, {Glanzman}, {Green}, {Grenier}, {Griffin}, {Grondin}, {Grove},
  {Guiriec}, {Harding}, {Hayashi}, {Hays}, {Hewitt}, {Horan},
  {J{\'o}hannesson}, {Johnson}, {Kamae}, {Kerr}, {Kocevski}, {Kovac'evic'},
  {Kuss}, {Landriu}, {Larsson}, {Latronico}, {Lemoine-Goumard}, {Li},
  {Liodakis}, {Longo}, {Loparco}, {Lott}, {Lovellette}, {Lubrano}, {Madejski},
  {Maldera}, {Malyshev}, {Manfreda}, {Marchesini}, {Marcotulli},
  {Mart{\'\i}-Devesa}, {Martin}, {Massaro}, {Mazziotta}, {McEnery}, {Mereu},
  {Meyer}, {Michelson}, {Mirabal}, {Mizuno}, {Monzani}, {Morselli},
  {Moskalenko}, {Negro}, {Nuss}, {Ojha}, {Omodei}, {Orienti}, {Orlando},
  {Ormes}, {Palatiello}, {Paliya}, {Paneque}, {Pei}, {Pe{\~n}a-Herazo},
  {Perkins}, {Persic}, {Pesce-Rollins}, {Petrosian}, {Petrov}, {Piron}, {Poon},
  {Porter}, {Principe}, {Rain{\`o}}, {Rando}, {Razzano}, {Razzaque}, {Reimer},
  {Reimer}, {Remy}, {Reposeur}, {Romani}, {Saz Parkinson}, {Schinzel},
  {Serini}, {Sgr{\`o}}, {Siskind}, {Smith}, {Spandre}, {Spinelli}, {Strong},
  {Suson}, {Tajima}, {Takahashi}, {Tak}, {Thayer}, {Thompson}, {Tibaldo},
  {Torres}, {Torresi}, {Valverde}, {Van Klaveren}, {van Zyl}, {Wood},
  {Yassine}, \& {Zaharijas}}]{4FGL}
{Abdollahi}, S., {Acero}, F., {Ackermann}, M., {et~al.} 2020, \apjs, 247, 33

\bibitem[{{Acero} {et~al.}(2010){Acero}, {Aharonian}, {Akhperjanian}, {Anton},
  {Barres de Almeida}, {Bazer-Bachi}, {Becherini}, {Behera}, {Bernl{\"o}hr},
  {Bochow}, {Boisson}, {Bolmont}, {Borrel}, {Braun}, {Brucker}, {Brun}, {Brun},
  {B{\"u}hler}, {Bulik}, {B{\"u}sching}, {Boutelier}, {Chadwick},
  {Charbonnier}, {Chaves}, {Cheesebrough}, {Conrad}, {Chounet}, {Clapson},
  {Coignet}, {Dalton}, {Daniel}, {Davids}, {Degrange}, {Deil}, {Dickinson},
  {Djannati-Ata{\"\i}}, {Domainko}, {Drury}, {Dubois}, {Dubus}, {Dyks},
  {Dyrda}, {Egberts}, {Eger}, {Espigat}, {Fallon}, {Farnier}, {Fegan},
  {Feinstein}, {Fiasson}, {F{\"o}rster}, {Fontaine}, {F{\"u}{\ss}ling},
  {Gabici}, {Gallant}, {G{\'e}rard}, {Gerbig}, {Giebels}, {Glicenstein},
  {Gl{\"u}ck}, {Goret}, {G{\"o}ring}, {Hauser}, {Heinz}, {Heinzelmann},
  {Henri}, {Hermann}, {Hinton}, {Hoffmann}, {Hofmann}, {Hofverberg},
  {Holleran}, {Hoppe}, {Horns}, {Jacholkowska}, {de Jager}, {Jahn}, {Jung},
  {Katarzy{\'n}ski}, {Katz}, {Kaufmann}, {Kerschhaggl}, {Khangulyan},
  {Kh{\'e}lifi}, {Keogh}, {Klochkov}, {Klu{\'z}niak}, {Kneiske}, {Komin},
  {Kosack}, {Kossakowski}, {Lamanna}, {Lenain}, {Lohse}, {Marandon},
  {Martineau-Huynh}, {Marcowith}, {Masbou}, {Maurin}, {McComb}, {Medina},
  {M{\'e}hault}, {Moderski}, {Moulin}, {Naumann-Godo}, {de Naurois}, {Nedbal},
  {Nekrassov}, {Nicholas}, {Niemiec}, {Nolan}, {Ohm}, {Olive}, {de O{\~n}a
  Wilhelmi}, {Orford}, {Ostrowski}, {Panter}, {Arribas}, {Pedaletti},
  {Pelletier}, {Petrucci}, {Pita}, {P{\"u}hlhofer}, {Punch}, {Quirrenbach},
  {Raubenheimer}, {Raue}, {Rayner}, {Reimer}, {Renaud}, {Rieger}, {Ripken},
  {Rob}, {Rosier-Lees}, {Rowell}, {Rudak}, {Rulten}, {Ruppel}, {Ryde},
  {Sahakian}, {Santangelo}, {Schlickeiser}, {Sch{\"o}ck}, {Sch{\"o}nwald},
  {Schwanke}, {Schwarzburg}, {Schwemmer}, {Shalchi}, {Sikora}, {Skilton},
  {Sol}, {Stawarz}, {Steenkamp}, {Stegmann}, {Stinzing}, {Superina}, {Sushch},
  {Szostek}, {Tam}, {Tavernet}, {Terrier}, {Tibolla}, {Tluczykont}, {van
  Eldik}, {Vasileiadis}, {Venter}, {Venter}, {Vialle}, {Vincent}, {Vivier},
  {V{\"o}lk}, {Volpe}, {Wagner}, {Ward}, {Zdziarski}, {Zech}, \&
  {H.\,E.\,S.\,S. Collaboration}}]{pointing}
{Acero}, F., {Aharonian}, F., {Akhperjanian}, A.~G., {et~al.} 2010, \mnras,
  402, 1877

\bibitem[{{Aharonian} {et~al.}(2005){Aharonian}, {Akhperjanian}, {Aye},
  {Bazer-Bachi}, {Beilicke}, {Benbow}, {Berge}, {Berghaus}, {Bernl{\"o}hr},
  {Boisson}, {Bolz}, {Borgmeier}, {Braun}, {Breitling}, {Brown}, {Gordo},
  {Chadwick}, {Chounet}, {Cornils}, {Costamante}, {Degrange},
  {Djannati-Ata{\"\i}}, {Drury}, {Dubus}, {Ergin}, {Espigat}, {Feinstein},
  {Fleury}, {Fontaine}, {Funk}, {Gallant}, {Giebels}, {Gillessen}, {Goret},
  {Hadjichristidis}, {Hauser}, {Heinzelmann}, {Henri}, {Hermann}, {Hinton},
  {Hofmann}, {Holleran}, {Horns}, {de Jager}, {Jung}, {Kh{\'e}lifi}, {Komin},
  {Konopelko}, {Latham}, {Le Gallou}, {Lemi{\`e}re}, {Lemoine}, {Leroy},
  {Lohse}, {Marcowith}, {Masterson}, {McComb}, {de Naurois}, {Nolan},
  {Noutsos}, {Orford}, {Osborne}, {Ouchrif}, {Panter}, {Pelletier}, {Pita},
  {P{\"u}hlhofer}, {Punch}, {Raubenheimer}, {Raue}, {Raux}, {Rayner},
  {Redondo}, {Reimer}, {Reimer}, {Ripken}, {Rob}, {Rolland}, {Rowell},
  {Sahakian}, {Saug{\'e}}, {Schlenker}, {Schlickeiser}, {Schuster}, {Schwanke},
  {Siewert}, {Sol}, {Steenkamp}, {Stegmann}, {Tavernet}, {Terrier},
  {Th{\'e}oret}, {Tluczykont}, {van der Walt}, {Vasileiadis}, {Venter},
  {Vincent}, {Visser}, {V{\"o}lk}, \& {Wagner}}]{firstHGPS}
{Aharonian}, F., {Akhperjanian}, A.~G., {Aye}, K.~M., {et~al.} 2005, Science,
  307, 1938

\bibitem[{{Aharonian} {et~al.}(2006{\natexlab{a}}){Aharonian}, {Akhperjanian},
  {Bazer-Bachi}, {Beilicke}, {Benbow}, {Berge}, {Bernl{\"o}hr}, {Boisson},
  {Bolz}, {Borrel}, {Braun}, {Breitling}, {Brown}, {B{\"u}hler},
  {B{\"u}sching}, {Carrigan}, {Chadwick}, {Chounet}, {Cornils}, {Costamante},
  {Degrange}, {Dickinson}, {Djannati-Ata{\"\i}}, {O'C. Drury}, {Dubus},
  {Egberts}, {Emmanoulopoulos}, {Espigat}, {Feinstein}, {Ferrero}, {Fiasson},
  {Fontaine}, {Funk}, {Funk}, {Gallant}, {Giebels}, {Glicenstein}, {Goret},
  {Hadjichristidis}, {Hauser}, {Hauser}, {Heinzelmann}, {Henri}, {Hermann},
  {Hinton}, {Hofmann}, {Holleran}, {Horns}, {Jacholkowska}, {de Jager},
  {Kh{\'e}lifi}, {Komin}, {Konopelko}, {Kosack}, {Latham}, {Le Gallou},
  {Lemi{\`e}re}, {Lemoine-Goumard}, {Lohse}, {Martin}, {Martineau-Huynh},
  {Marcowith}, {Masterson}, {McComb}, {de Naurois}, {Nedbal}, {Nolan},
  {Noutsos}, {Orford}, {Osborne}, {Ouchrif}, {Panter}, {Pelletier}, {Pita},
  {P{\"u}hlhofer}, {Punch}, {Raubenheimer}, {Raue}, {Rayner}, {Reimer},
  {Reimer}, {Ripken}, {Rob}, {Rolland}, {Rowell}, {Sahakian}, {Saug{\'e}},
  {Schlenker}, {Schlickeiser}, {Schwanke}, {Sol}, {Spangler}, {Spanier},
  {Steenkamp}, {Stegmann}, {Superina}, {Tavernet}, {Terrier}, {Th{\'e}oret},
  {Tluczykont}, {van Eldik}, {Vasileiadis}, {Venter}, {Vincent}, {V{\"o}lk},
  {Wagner}, \& {Ward}}]{sys_index}
{Aharonian}, F., {Akhperjanian}, A.~G., {Bazer-Bachi}, A.~R., {et~al.}
  2006{\natexlab{a}}, \aap, 457, 899

\bibitem[{{Aharonian} {et~al.}(2006{\natexlab{b}}){Aharonian}, {Akhperjanian},
  {Bazer-Bachi}, {Beilicke}, {Benbow}, {Berge}, {Bernl{\"o}hr}, {Boisson},
  {Bolz}, {Borrel}, {Braun}, {Breitling}, {Brown}, {Chadwick}, {Chounet},
  {Cornils}, {Costamante}, {Degrange}, {Dickinson}, {Djannati-Ata{\"\i}},
  {Drury}, {Dubus}, {Emmanoulopoulos}, {Espigat}, {Feinstein}, {Fontaine},
  {Fuchs}, {Funk}, {Gallant}, {Giebels}, {Gillessen}, {Glicenstein}, {Goret},
  {Hadjichristidis}, {Hauser}, {Heinzelmann}, {Henri}, {Hermann}, {Hinton},
  {Hofmann}, {Holleran}, {Horns}, {Jacholkowska}, {de Jager}, {Kh{\'e}lifi},
  {Komin}, {Konopelko}, {Latham}, {Le Gallou}, {Lemi{\`e}re},
  {Lemoine-Goumard}, {Leroy}, {Lohse}, {Martin}, {Martineau-Huynh},
  {Marcowith}, {Masterson}, {McComb}, {de Naurois}, {Nolan}, {Noutsos},
  {Orford}, {Osborne}, {Ouchrif}, {Panter}, {Pelletier}, {Pita},
  {P{\"u}hlhofer}, {Punch}, {Raubenheimer}, {Raue}, {Raux}, {Rayner}, {Reimer},
  {Reimer}, {Ripken}, {Rob}, {Rolland}, {Rowell}, {Sahakian}, {Saug{\'e}},
  {Schlenker}, {Schlickeiser}, {Schuster}, {Schwanke}, {Siewert}, {Sol},
  {Spangler}, {Steenkamp}, {Stegmann}, {Tavernet}, {Terrier}, {Th{\'e}oret},
  {Tluczykont}, {Vasileiadis}, {Venter}, {Vincent}, {V{\"o}lk}, \&
  {Wagner}}]{HGPS}
{Aharonian}, F., {Akhperjanian}, A.~G., {Bazer-Bachi}, A.~R., {et~al.}
  2006{\natexlab{b}}, \apj, 636, 777

\bibitem[{{Ajello} {et~al.}(2021){Ajello}, {Atwood}, {Axelsson}, {Bagagli},
  {Bagni}, {Baldini}, {Bastieri}, {Bellardi}, {Bellazzini}, {Bissaldi},
  {Bloom}, {Bonino}, {Bregeon}, {Brez}, {Bruel}, {Buehler}, {Buson}, {Cameron},
  {Caraveo}, {Cavazzuti}, {Ceccanti}, {Chen}, {Cheung}, {Ciprini}, {Cognard},
  {Cohen-Tanugi}, {Cutini}, {D'Ammando}, {de la Torre Luque}, {de Palma},
  {Digel}, {Dirirsa}, {Di Lalla}, {Di Venere}, {Dom{\'\i}nguez}, {Fabiani},
  {Ferrara}, {Fiori}, {Foglia}, {Fukazawa}, {Fusco}, {Gargano}, {Gasparrini},
  {Giroletti}, {Glanzman}, {Green}, {Griffin}, {Grondin}, {Grove}, {Guillemot},
  {Guiriec}, {Gustafsson}, {Hays}, {Horan}, {J{\'o}hannesson}, {Johnson},
  {Kamae}, {Kerr}, {Kuss}, {Larsson}, {Latronico}, {Lemoine-Goumard}, {Li},
  {Liodakis}, {Longo}, {Loparco}, {Lovellette}, {Lubrano}, {Maldera},
  {Manfreda}, {Mart{\'\i}-Devesa}, {Mazziotta}, {Menon}, {Mereu}, {Meyer},
  {Michelson}, {Minuti}, {Mitthumsiri}, {Mizuno}, {Mongelli}, {Monzani},
  {Moskalenko}, {Negro}, {Nuss}, {Ojha}, {Orienti}, {Orlando}, {Paccagnella},
  {Paliya}, {Paneque}, {Pei}, {Perkins}, {Pesce-Rollins}, {Pinchera}, {Piron},
  {Poon}, {Porter}, {Primavera}, {Principe}, {Racusin}, {Rain{\`o}}, {Rando},
  {Rani}, {Rapposelli}, {Razzano}, {Razzaque}, {Reimer}, {Reimer}, {Russell},
  {Saggini}, {Saz Parkinson}, {Scolieri}, {Serini}, {Sgr{\`o}}, {Siskind},
  {Smith}, {Spandre}, {Spinelli}, {Suson}, {Tajima}, {Thayer}, {Thompson},
  {Tibaldo}, {Torres}, {Tosti}, {Valverde}, {Vigiani}, \&
  {Zaharijas}}]{sensitivity}
{Ajello}, M., {Atwood}, W.~B., {Axelsson}, M., {et~al.} 2021, \apjs, 256, 12

\bibitem[{{Albert} {et~al.}(2020){Albert}, {Alfaro}, {Alvarez}, {Camacho},
  {Arteaga-Vel{\'a}zquez}, {Arunbabu}, {Avila Rojas}, {Ayala Solares},
  {Baghmanyan}, {Belmont-Moreno}, {BenZvi}, {Brisbois}, {Caballero-Mora},
  {Capistr{\'a}n}, {Carrami{\~n}ana}, {Casanova}, {Cotti}, {Couti{\~n}o de
  Le{\'o}n}, {De la Fuente}, {Diaz Hernandez}, {Diaz-Cruz}, {Dingus},
  {DuVernois}, {Durocher}, {D{\'\i}az-V{\'e}lez}, {Ellsworth}, {Engel},
  {Espinoza}, {Fan}, {Fang}, {Alonso}, {Fleischhack}, {Fraija},
  {Galv{\'a}n-G{\'a}mez}, {Garcia}, {Garc{\'\i}a-Gonz{\'a}lez}, {Garfias},
  {Giacinti}, {Gonz{\'a}lez}, {Goodman}, {Harding}, {Hernandez}, {Hinton},
  {Hona}, {Huang}, {Hueyotl-Zahuantitla}, {H{\"u}ntemeyer}, {Iriarte},
  {Jardin-Blicq}, {Joshi}, {Kieda}, {Lara}, {Lee}, {Le{\'o}n Vargas},
  {Linnemann}, {Longinotti}, {Luis-Raya}, {Lundeen}, {L{\'o}pez-Coto},
  {Malone}, {Marandon}, {Martinez}, {Martinez-Castellanos},
  {Mart{\'\i}nez-Castro}, {Matthews}, {Miranda-Romagnoli}, {Morales-Soto},
  {Moreno}, {Mostaf{\'a}}, {Nayerhoda}, {Nellen}, {Newbold}, {Nisa},
  {Noriega-Papaqui}, {Olivera-Nieto}, {Omodei}, {Peisker}, {P{\'e}rez Araujo},
  {P{\'e}rez-P{\'e}rez}, {Ren}, {Rho}, {Rivi{\`e}re}, {Rosa-Gonz{\'a}lez},
  {Ruiz-Velasco}, {Salazar}, {Salesa Greus}, {Sandoval}, {Schneider},
  {Schoorlemmer}, {Serna}, {Sinnis}, {Smith}, {Springer}, {Surajbali},
  {Tollefson}, {Torres}, {Torres-Escobedo}, {Ukwatta}, {Ure{\~n}a-Mena},
  {Weisgarber}, {Werner}, {Willox}, {Zepeda}, {Zhou}, {de Le{\'o}n},
  {{\'A}lvarez}, \& {HAWC Collaboration}}]{hawc_cat}
{Albert}, A., {Alfaro}, R., {Alvarez}, C., {et~al.} 2020, \apj, 905, 76

\bibitem[{{Albert} {et~al.}(2006){Albert}, {Aliu}, {Anderhub}, {Antoranz},
  {Armada}, {Asensio}, {Baixeras}, {Barrio}, {Bartel}, {Bartko}, {Bastieri},
  {Bavikadi}, {Bednarek}, {Berger}, {Bigongiari}, {Biland}, {Bisesi}, {Blanch},
  {Bock}, {Bretz}, {Britvitch}, {Camara}, {Chilingarian}, {Ciprini}, {Coarasa},
  {Commichau}, {Contreras}, {Cortina}, {Curtev}, {Danielyan}, {Dazzi}, {De
  Angelis}, {de los Reyes}, {De Lotto}, {Domingo-Santamaria}, {Dorner}, {Doro},
  {Errando}, {Fagiolini}, {Ferenc}, {Fern{\'a}ndez}, {Firpo}, {Flix},
  {Fonseca}, {Font}, {Galante}, {Garczarczyk}, {Gaug}, {Gebauer}, {Giller},
  {Goebel}, {Hakobyan}, {Hayashida}, {Hengstebeck}, {H{\"o}hne}, {Hose},
  {Jacon}, {Kalekin}, {Kranich}, {Laille}, {Lenisa}, {Liebing}, {Lindfors},
  {Longo}, {L{\'o}pez}, {L{\'o}pez}, {Lorenz}, {Lucarelli}, {Majumdar},
  {Maneva}, {Mannheim}, {Mariotti}, {Mart{\'\i}nez}, {Mase}, {Mazin}, {Merck},
  {Merck}, {Meucci}, {Meyer}, {Miranda}, {Mirzoyan}, {Mizobuchi}, {Moralejo},
  {Nilsson}, {O{\~n}a-Wilhelmi}, {Ordu{\~n}a}, {Otte}, {Oya}, {Paneque},
  {Paoletti}, {Pasanen}, {Pascoli}, {Pauss}, {Pavel}, {Pegna}, {Peruzzo},
  {Piccioli}, {Prandini}, {Rico}, {Rhode}, {Riegel}, {Rissi}, {Robert},
  {Rossato}, {R{\"u}gamer}, {Saggion}, {Sanchez}, {Sartori}, {Scalzotto},
  {Schmitt}, {Schweizer}, {Shayduk}, {Shinozaki}, {Shore}, {Sidro},
  {Sillanp{\"a}{\"a}}, {Sobczynska}, {Stamerra}, {Stark}, {Takalo}, {Temnikov},
  {Tescaro}, {Teshima}, {Tonello}, {Torres}, {Torres}, {Turini}, {Vankov},
  {Vitale}, {Wagner}, {Wibig}, {Wittek}, \& {Zapatero}}]{magic}
{Albert}, J., {Aliu}, E., {Anderhub}, H., {et~al.} 2006, \apjl, 637, L41

\bibitem[{{Araya}(2018)}]{fermi_araya}
{Araya}, M. 2018, \apj, 859, 69

\bibitem[{{Atwood} {et~al.}(2009){Atwood}, {Abdo}, {Ackermann}, {Althouse},
  {Anderson}, {Axelsson}, {Baldini}, {Ballet}, {Band}, {Barbiellini},
  {Bartelt}, {Bastieri}, {Baughman}, {Bechtol}, {B{\'e}d{\'e}r{\`e}de},
  {Bellardi}, {Bellazzini}, {Berenji}, {Bignami}, {Bisello}, {Bissaldi},
  {Blandford}, {Bloom}, {Bogart}, {Bonamente}, {Bonnell}, {Borgland},
  {Bouvier}, {Bregeon}, {Brez}, {Brigida}, {Bruel}, {Burnett}, {Busetto},
  {Caliandro}, {Cameron}, {Caraveo}, {Carius}, {Carlson}, {Casandjian},
  {Cavazzuti}, {Ceccanti}, {Cecchi}, {Charles}, {Chekhtman}, {Cheung},
  {Chiang}, {Chipaux}, {Cillis}, {Ciprini}, {Claus}, {Cohen-Tanugi},
  {Condamoor}, {Conrad}, {Corbet}, {Corucci}, {Costamante}, {Cutini}, {Davis},
  {Decotigny}, {DeKlotz}, {Dermer}, {de Angelis}, {Digel}, {do Couto e Silva},
  {Drell}, {Dubois}, {Dumora}, {Edmonds}, {Fabiani}, {Farnier}, {Favuzzi},
  {Flath}, {Fleury}, {Focke}, {Funk}, {Fusco}, {Gargano}, {Gasparrini},
  {Gehrels}, {Gentit}, {Germani}, {Giebels}, {Giglietto}, {Giommi}, {Giordano},
  {Glanzman}, {Godfrey}, {Grenier}, {Grondin}, {Grove}, {Guillemot}, {Guiriec},
  {Haller}, {Harding}, {Hart}, {Hays}, {Healey}, {Hirayama}, {Hjalmarsdotter},
  {Horn}, {Hughes}, {J{\'o}hannesson}, {Johansson}, {Johnson}, {Johnson},
  {Johnson}, {Johnson}, {Kamae}, {Katagiri}, {Kataoka}, {Kavelaars}, {Kawai},
  {Kelly}, {Kerr}, {Klamra}, {Kn{\"o}dlseder}, {Kocian}, {Komin}, {Kuehn},
  {Kuss}, {Landriu}, {Latronico}, {Lee}, {Lee}, {Lemoine-Goumard}, {Lionetto},
  {Longo}, {Loparco}, {Lott}, {Lovellette}, {Lubrano}, {Madejski}, {Makeev},
  {Marangelli}, {Massai}, {Mazziotta}, {McEnery}, {Menon}, {Meurer},
  {Michelson}, {Minuti}, {Mirizzi}, {Mitthumsiri}, {Mizuno}, {Moiseev},
  {Monte}, {Monzani}, {Moretti}, {Morselli}, {Moskalenko}, {Murgia},
  {Nakamori}, {Nishino}, {Nolan}, {Norris}, {Nuss}, {Ohno}, {Ohsugi}, {Omodei},
  {Orlando}, {Ormes}, {Paccagnella}, {Paneque}, {Panetta}, {Parent}, {Pearce},
  {Pepe}, {Perazzo}, {Pesce-Rollins}, {Picozza}, {Pieri}, {Pinchera}, {Piron},
  {Porter}, {Poupard}, {Rain{\`o}}, {Rando}, {Rapposelli}, {Razzano}, {Reimer},
  {Reimer}, {Reposeur}, {Reyes}, {Ritz}, {Rochester}, {Rodriguez}, {Romani},
  {Roth}, {Russell}, {Ryde}, {Sabatini}, {Sadrozinski}, {Sanchez}, {Sander},
  {Sapozhnikov}, {Parkinson}, {Scargle}, {Schalk}, {Scolieri}, {Sgr{\`o}},
  {Share}, {Shaw}, {Shimokawabe}, {Shrader}, {Sierpowska-Bartosik}, {Siskind},
  {Smith}, {Smith}, {Spandre}, {Spinelli}, {Starck}, {Stephens}, {Strickman},
  {Strong}, {Suson}, {Tajima}, {Takahashi}, {Takahashi}, {Tanaka}, {Tenze},
  {Tether}, {Thayer}, {Thayer}, {Thompson}, {Tibaldo}, {Tibolla}, {Torres},
  {Tosti}, {Tramacere}, {Turri}, {Usher}, {Vilchez}, {Vitale}, {Wang},
  {Watters}, {Winer}, {Wood}, {Ylinen}, \& {Ziegler}}]{LAT_performance}
{Atwood}, W.~B., {Abdo}, A.~A., {Ackermann}, M., {et~al.} 2009, \apj, 697, 1071

\bibitem[{{Bernl{\"o}hr}(2008)}]{IRFs}
{Bernl{\"o}hr}, K. 2008, Astroparticle Physics, 30, 149

\bibitem[{{Brogan} {et~al.}(2005{\natexlab{a}}){Brogan}, {Gaensler}, {Gelfand},
  {Lazendic}, {Lazio}, {Kassim}, \& {McClure-Griffiths}}]{w33}
{Brogan}, C.~L., {Gaensler}, B.~M., {Gelfand}, J.~D., {et~al.}
  2005{\natexlab{a}}, \apjl, 629, L105

\bibitem[{{Brogan} {et~al.}(2005{\natexlab{b}}){Brogan}, {Gaensler}, {Gelfand},
  {Lazendic}, {Lazio}, {Kassim}, \& {McClure-Griffiths}}]{VLA}
{Brogan}, C.~L., {Gaensler}, B.~M., {Gelfand}, J.~D., {et~al.}
  2005{\natexlab{b}}, \apjl, 629, L105

\bibitem[{{Camilo} {et~al.}(2021){Camilo}, {Ransom}, {Halpern}, \&
  {Roshi}}]{psr_distance1}
{Camilo}, F., {Ransom}, S.~M., {Halpern}, J.~P., \& {Roshi}, D.~A. 2021, \apj,
  917, 67

\bibitem[{{Cao} {et~al.}(2023){Cao}, {Aharonian}, {An}, {Axikegu}, {Bai},
  {Bao}, {Bastieri}, {Bi}, {Bi}, {Cai}, {Cao}, {Cao}, {Cao}, {Chang}, {Chang},
  {Chen}, {Chen}, {Chen}, {Chen}, {Chen}, {Chen}, {Chen}, {Chen}, {Chen},
  {Chen}, {Chen}, {Chen}, {Cheng}, {Cheng}, {Cui}, {Cui}, {Cui}, {Cui}, {Dai},
  {Dai}, {Dai}, {Danzengluobu}, {della Volpe}, {Dong}, {Duan}, {Fan}, {Fan},
  {Fang}, {Fang}, {Feng}, {Feng}, {Feng}, {Feng}, {Feng}, {Gabici}, {Gao},
  {Gao}, {Gao}, {Gao}, {Gao}, {Gao}, {Ge}, {Geng}, {Giacinti}, {Gong}, {Gou},
  {Gu}, {Guo}, {Guo}, {Guo}, {Guo}, {Han}, {He}, {He}, {He}, {He}, {He},
  {Heller}, {Hor}, {Hou}, {Hou}, {Hou}, {Hu}, {Hu}, {Hu}, {Huang}, {Huang},
  {Huang}, {Huang}, {Huang}, {Huang}, {Huang}, {Ji}, {Jia}, {Jia}, {Jiang},
  {Jiang}, {Jiang}, {Jin}, {Kang}, {Ke}, {Kuleshov}, {Kurinov}, {Li}, {Li},
  {Li}, {Li}, {Li}, {Li}, {Li}, {Li}, {Li}, {Li}, {Li}, {Li}, {Li}, {Li}, {Li},
  {Li}, {Li}, {Li}, {Li}, {Liang}, {Liang}, {Lin}, {Liu}, {Liu}, {Liu}, {Liu},
  {Liu}, {Liu}, {Liu}, {Liu}, {Liu}, {Liu}, {Liu}, {Liu}, {Liu}, {Liu}, {Lu},
  {Luo}, {Lv}, {Ma}, {Ma}, {Ma}, {Mao}, {Min}, {Mitthumsiri}, {Mu}, {Nan},
  {Neronov}, {Ou}, {Pang}, {Pattarakijwanich}, {Pei}, {Qi}, {Qi}, {Qiao},
  {Qin}, {Ruffolo}, {S{\'a}iz}, {Semikoz}, {Shao}, {Shao}, {Shchegolev},
  {Sheng}, {Shu}, {Song}, {Stenkin}, {Stepanov}, {Su}, {Sun}, {Sun}, {Sun},
  {Tam}, {Tang}, {Tang}, {Tian}, {Wang}, {Wang}, {Wang}, {Wang}, {Wang},
  {Wang}, {Wang}, {Wang}, {Wang}, {Wang}, {Wang}, {Wang}, {Wang}, {Wang},
  {Wang}, {Wang}, {Wang}, {Wang}, {Wang}, {Wang}, {Wang}, {Wei}, {Wei}, {Wei},
  {Wen}, {Wu}, {Wu}, {Wu}, {Wu}, {Wu}, {Xi}, {Xia}, {Xia}, {Xiang}, {Xiao},
  {Xiao}, {Xin}, {Xin}, {Xing}, {Xiong}, {Xu}, {Xu}, {Xu}, {Xu}, {Xue}, {Yan},
  {Yan}, {Yan}, {Yang}, {Yang}, {Yang}, {Yang}, {Yang}, {Yang}, {Yang}, {Yang},
  {Yang}, {Yao}, {Yao}, {Ye}, {Yin}, {Yin}, {You}, {You}, {Yu}, {Yuan}, {Yue},
  {Zeng}, {Zeng}, {Zeng}, {Zha}, {Zhang}, {Zhang}, {Zhang}, {Zhang}, {Zhang},
  {Zhang}, {Zhang}, {Zhang}, {Zhang}, {Zhang}, {Zhang}, {Zhang}, {Zhang},
  {Zhang}, {Zhang}, {Zhang}, {Zhang}, {Zhang}, {Zhao}, {Zhao}, {Zhao}, {Zhao},
  {Zhao}, {Zheng}, {Zhou}, {Zhou}, {Zhou}, {Zhou}, {Zhou}, {Zhou}, {Zhou},
  {Zhu}, {Zhu}, {Zhu}, {Zhu}, \& {Zuo.}}]{lhaaso}
{Cao}, Z., {Aharonian}, F., {An}, Q., {et~al.} 2023, arXiv e-prints,
  arXiv:2305.17030

\bibitem[{Deil {et~al.}(2020)Deil, Donath, Terrier, Ruiz, King, QRemy, Sinha,
  Wood, fabiopintore, mapazarr, LauraOlivera, luca giunti, ellisowen, Sipőcz,
  Vorokh, Robitaille, JonathanDHarris, Mohrmann, Khelifi, Jaleleddine, Nigro,
  Chakraborty, Droettboom, china\_108, Watson, Voruganti, Tollerud,
  thomasarmstrong, \& Tiziani}]{gammapy_zenodo}
Deil, C., Donath, A., Terrier, R., {et~al.} 2020, gammapy/gammapy: v.0.18.2

\bibitem[{{Diesing} \& {Caprioli}(2019)}]{hadron_injection}
{Diesing}, R. \& {Caprioli}, D. 2019, \prl, 123, 071101

\bibitem[{{Donath} {et~al.}(2023){Donath}, {Terrier}, {Remy}, {Sinha}, {Nigro},
  {Pintore}, {Kh{\'e}lifi}, {Olivera-Nieto}, {Ruiz}, {Br{\"u}gge}, {Linhoff},
  {Contreras}, {Acero}, {Aguasca-Cabot}, {Berge}, {Bhattacharjee}, {Buchner},
  {Boisson}, {Carreto Fidalgo}, {Chen}, {de Bony de Lavergne}, {Cardoso},
  {Deil}, {F{\"u}{\ss}ling}, {Funk}, {Giunti}, {Hinton}, {Jouvin}, {King},
  {Lefaucheur}, {Lemoine-Goumard}, {Lenain}, {L{\'o}pez-Coto}, {Mohrmann},
  {Morcuende}, {Panny}, {Regeard}, {Saha}, {Siejkowski}, {Siemiginowska},
  {Sip{\H{o}}cz}, {Unbehaun}, {van Eldik}, {Vuillaume}, \& {Zanin}}]{gammapy}
{Donath}, A., {Terrier}, R., {Remy}, Q., {et~al.} 2023, arXiv e-prints,
  arXiv:2308.13584

\bibitem[{{Dzib} \& {Rodr{\'\i}guez}(2021)}]{quasars}
{Dzib}, S.~A. \& {Rodr{\'\i}guez}, L.~F. 2021, \apj, 923, 228

\bibitem[{\emph{Fermi} LAT~Collaboration(2019)}]{fermi_sys}
\emph{Fermi} LAT~Collaboration. 2019, Propagating the Uncertainties on the
  Effective Area (Aeff), [Online; accessed 12-September-2023]

\bibitem[{{Foreman-Mackey} {et~al.}(2013){Foreman-Mackey}, {Hogg}, {Lang}, \&
  {Goodman}}]{emcee}
{Foreman-Mackey}, D., {Hogg}, D.~W., {Lang}, D., \& {Goodman}, J. 2013, \pasp,
  125, 306

\bibitem[{{Funk} {et~al.}(2007){Funk}, {Hinton}, {Moriguchi}, {Aharonian},
  {Fukui}, {Hofmann}, {Horns}, {P{\"u}hlhofer}, {Reimer}, {Rowell}, {Terrier},
  {Vink}, \& {Wagner}}]{xmm-newton}
{Funk}, S., {Hinton}, J.~A., {Moriguchi}, Y., {et~al.} 2007, \aap, 470, 249

\bibitem[{{Gaensler} \& {Slane}(2006)}]{modeling2}
{Gaensler}, B.~M. \& {Slane}, P.~O. 2006, \araa, 44, 17

\bibitem[{{Giacinti} {et~al.}(2020){Giacinti}, {Mitchell}, {L{\'o}pez-Coto},
  {Joshi}, {Parsons}, \& {Hinton}}]{pwn_population}
{Giacinti}, G., {Mitchell}, A.~M.~W., {L{\'o}pez-Coto}, R., {et~al.} 2020,
  \aap, 636, A113

\bibitem[{{Gotthelf} \& {Halpern}(2009)}]{psr}
{Gotthelf}, E.~V. \& {Halpern}, J.~P. 2009, \apjl, 700, L158

\bibitem[{{Hahn}(2015)}]{gamera}
{Hahn}, J. 2015, in 34th International Cosmic Ray Conference (ICRC2015),
  Vol.~34, 917

\bibitem[{{Helfand} {et~al.}(2007){Helfand}, {Gotthelf}, {Halpern}, {Camilo},
  {Semler}, {Becker}, \& {White}}]{chandra}
{Helfand}, D.~J., {Gotthelf}, E.~V., {Halpern}, J.~P., {et~al.} 2007, \apj,
  665, 1297

\bibitem[{{H.E.S.S. Collaboration}(2018)}]{HGPS2}
{H.E.S.S. Collaboration}. 2018, \aap, 612, A1

\bibitem[{{H.E.S.S. Collaboration}(2019)}]{J1825}
{H.E.S.S. Collaboration}. 2019, \aap, 621, A116

\bibitem[{{H.E.S.S. Collaboration}(2023)}]{J1809}
{H.E.S.S. Collaboration}. 2023, \aap, 672, A103

\bibitem[{{H.\,E.\,S.\,S. Collaboration} {et~al.}(2023){H.\,E.\,S.\,S.
  Collaboration}, {Aharonian}, {Ait Benkhali}, {Aschersleben}, {Ashkar},
  {Backes}, {Barbosa Martins}, {Batzofin}, {Becherini}, {Berge},
  {B{\"o}ttcher}, {Boisson}, {Bolmont}, {Borowska}, {Bouyahiaoui}, {Bradascio},
  {Breuhaus}, {Brose}, {Brun}, {Bruno}, {Bulik}, {Burger-Scheidlin}, {Bylund},
  {Caroff}, {Casanova}, {Celic}, {Cerruti}, {Chambery}, {Chand}, {Chen},
  {Chibueze}, {Chibueze}, {Damascene Mbarubucyeye}, {Djannati-Ata{\"\i}},
  {Dmytriiev}, {Einecke}, {Ernenwein}, {Feijen}, {Filipovic}, {Fontaine},
  {F{\"u}{\ss}ling}, {Funk}, {Gabici}, {Gallant}, {Ghafourizadeh}, {Giavitto},
  {Giunti}, {Glawion}, {Goswami}, {Grolleron}, {Grondin}, {Haerer}, {Hinton},
  {Hofmann}, {Holch}, {Holler}, {Horns}, {Huang}, {Jamrozy}, {Jankowsky},
  {Joshi}, {Jung-Richardt}, {Kasai}, {Katarzy{\'n}ski}, {Kh{\'e}lifi},
  {Klu{\'z}niak}, {Komin}, {Kosack}, {Kostunin}, {Lang}, {Le Stum}, {Leitl},
  {Lemi{\`e}re}, {Lemoine-Goumard}, {Lenain}, {Leuschner}, {Lohse},
  {Luashvili}, {Lypova}, {Mackey}, {Malyshev}, {Malyshev}, {Marandon},
  {Marchegiani}, {Marcowith}, {Marinos}, {Mart{\'\i}-Devesa}, {Marx},
  {Mitchell}, {Moderski}, {Mohrmann}, {Montanari}, {Moulin}, {Muller},
  {Nakashima}, {de Naurois}, {Niemiec}, {Priyana Noel}, {Ohm}, {Olivera-Nieto},
  {de Ona Wilhelmi}, {Ostrowski}, {Panny}, {Panter}, {Parsons}, {Prokhorov},
  {P{\"u}hlhofer}, {Punch}, {Quirrenbach}, {Reichherzer}, {Reimer}, {Reimer},
  {Renaud}, {Reville}, {Rieger}, {Rowell}, {Rudak}, {Sahakian}, {Santangelo},
  {Sasaki}, {Schutte}, {Schwanke}, {Shapopi}, {Sol}, {Specovius}, {Spencer},
  {Stawarz}, {Steenkamp}, {Steinmassl}, {Sushch}, {Suzuki}, {Takahashi},
  {Tanaka}, {Terrier}, {Thorpe-Morgan}, {Tsirou}, {Tsuji}, {Uchiyama}, {van
  Eldik}, {Vecchi}, {Veh}, {Venter}, {Vink}, {Wach}, {Wagner}, {White},
  {Wierzcholska}, {Wong}, {Zacharias}, {Zargaryan}, {Zdziarski}, {Zech},
  {Zouari}, \& {{\.Z}ywucka}}]{halo2}
{H.\,E.\,S.\,S. Collaboration}, {Aharonian}, F., {Ait Benkhali}, F., {et~al.}
  2023, \aap, 672, A103

\bibitem[{{Khelifi} {et~al.}(2015){Khelifi}, {Djannati-Ata{\"\i}}, {Jouvin},
  {Lefaucheur}, {Lemiere}, {Pita}, {Tavernier}, \& {Terrier}}]{hap-fr}
{Khelifi}, B., {Djannati-Ata{\"\i}}, A., {Jouvin}, L., {et~al.} 2015, in 34th
  International Cosmic Ray Conference (ICRC2015), Vol.~34, 837

\bibitem[{{Kolmogorov}(1991)}]{kolmogorov}
{Kolmogorov}, A.~N. 1991, Proceedings of the Royal Society of London Series A,
  434, 9

\bibitem[{{Kraichnan}(1965)}]{kraichnan}
{Kraichnan}, R.~H. 1965, Physics of Fluids, 8, 1385

\bibitem[{{Li} \& {Ma}(1983)}]{lima}
{Li}, T.~P. \& {Ma}, Y.~Q. 1983, \apj, 272, 317

\bibitem[{{MAGIC Collaboration}(2020)}]{enedep2}
{MAGIC Collaboration}. 2020, \mnras, 497, 3734

\bibitem[{{Messineo} {et~al.}(2011){Messineo}, {Davies}, {Figer}, {Kudritzki},
  {Valenti}, {Trombley}, {Najarro}, \& {Rich}}]{stellar_cluster}
{Messineo}, M., {Davies}, B., {Figer}, D.~F., {et~al.} 2011, \apj, 733, 41

\bibitem[{{Miville-Desch{\^e}nes} {et~al.}(2017){Miville-Desch{\^e}nes},
  {Murray}, \& {Lee}}]{clouds1}
{Miville-Desch{\^e}nes}, M.-A., {Murray}, N., \& {Lee}, E.~J. 2017, \apj, 834,
  57

\bibitem[{{Mohrmann} {et~al.}(2019){Mohrmann}, {Specovius}, {Tiziani}, {Funk},
  {Malyshev}, {Nakashima}, \& {van Eldik}}]{fov-bkgmodel}
{Mohrmann}, L., {Specovius}, A., {Tiziani}, D., {et~al.} 2019, \aap, 632, A72

\bibitem[{{Morlino} {et~al.}(2021){Morlino}, {Blasi}, {Peretti}, \&
  {Cristofari}}]{2021MNRAS.504.6096Morlino}
{Morlino}, G., {Blasi}, P., {Peretti}, E., \& {Cristofari}, P. 2021, \mnras,
  504, 6096

\bibitem[{{Parsons} \& {Hinton}(2014)}]{impact}
{Parsons}, R.~D. \& {Hinton}, J.~A. 2014, Astroparticle Physics, 56, 26

\bibitem[{{Principe} {et~al.}(2020){Principe}, {Mitchell}, {Caroff}, {Hinton},
  {Parsons}, \& {Funk}}]{enedep1}
{Principe}, G., {Mitchell}, A.~M.~W., {Caroff}, S., {et~al.} 2020, \aap, 640,
  A76

\bibitem[{{Rice} {et~al.}(2016){Rice}, {Goodman}, {Bergin}, {Beaumont}, \&
  {Dame}}]{clouds2}
{Rice}, T.~S., {Goodman}, A.~A., {Bergin}, E.~A., {Beaumont}, C., \& {Dame},
  T.~M. 2016, \apj, 822, 52

\bibitem[{{Strong} \& {Moskalenko}(1998)}]{ISM_diff}
{Strong}, A.~W. \& {Moskalenko}, I.~V. 1998, \apj, 509, 212

\bibitem[{{Ubertini} {et~al.}(2005){Ubertini}, {Bassani}, {Malizia}, {Bazzano},
  {Bird}, {Dean}, {De Rosa}, {Lebrun}, {Moran}, {Renaud}, {Stephen}, {Terrier},
  \& {Walter}}]{integral}
{Ubertini}, P., {Bassani}, L., {Malizia}, A., {et~al.} 2005, \apjl, 629, L109

\bibitem[{{Venter} \& {de Jager}(2007)}]{model3}
{Venter}, C. \& {de Jager}, O.~C. 2007, in WE-Heraeus Seminar on Neutron Stars
  and Pulsars 40 years after the Discovery, ed. W.~{Becker} \& H.~H. {Huang},
  40

\bibitem[{{Wood} {et~al.}(2017){Wood}, {Caputo}, {Charles}, {Di Mauro},
  {Magill}, {Perkins}, \& {Fermi-LAT Collaboration}}]{fermipy}
{Wood}, M., {Caputo}, R., {Charles}, E., {et~al.} 2017, in 35th International
  Cosmic Ray Conference (ICRC2017), Vol. 301, 824

\bibitem[{Xin \& Guo(2021)}]{fermi_2021}
Xin, Y. \& Guo, X. 2021, in Proceedings of 37th International Cosmic Ray
  Conference {\textemdash} PoS(ICRC2021), Vol. 395, 625

\end{thebibliography}
